\documentclass[10pt]{iopart}
\usepackage{amsfonts}
\usepackage{graphicx}
\newcommand{\phd}{\phantom{\dagger}}

\def\vet{\mathbf}

\def\a{\alpha}
\def\b{\beta}
\def\g{\gamma}
\def\G{\Gamma}
\def\d{\delta}
\def\D{\Delta}

\def\eps{\varepsilon}

\def\t{\tau}

\def\w{\omega}

\def\l{\lambda}

\def\s{\sigma}

\begin{document}
\title[Simple models for the quantum shuttle]
{Simple models suffice for the single dot quantum shuttle}

\author{
 A Donarini$^{\dag\ddag}$, T Novotn\'{y}$^\star$$^+$,  and A-P Jauho$^{\dag}$}

\address{\dag~NanoDTU, MIC --  Department of Micro and Nanotechnology,
Technical University of Denmark (DTU), Building 345 East, DK --
2800 Kongens Lyngby, Denmark}

\address{\ddag~Institut I - Theoretische Physik Universit\" at
Regensburg D-93040 Regensburg, Germany}

\address{$^\star$~Nano-Science Center, University of Copenhagen,
Universitetsparken 5, DK -- 2100 Copenhagen {\O}, Denmark}

\address{$^+$~Department of Electronic Structures, Faculty of Mathematics and
Physics, Charles University, Ke Karlovu 5, 121 16 Prague, Czech
Republic}

\begin{abstract}
A quantum shuttle is an archetypical nanoelectromechanical device,
where the mechanical degree of freedom is quantized. Using a
full-scale numerical solution of the  generalized master equation
describing the shuttle, we have recently shown [Novotn\'{y} {\it
et al.}, Phys. Rev. Lett. {\bf 92}, 248302 (2004)] that for
certain limits of the shuttle parameters one can distinguish three
distinct charge transport mechanisms: (i) an incoherent tunneling
regime, (ii) a shuttling regime, where the charge transport is
synchronous with the mechanical motion, and (iii) a coexistence
regime, where the device switches between the tunneling and
shuttling regimes. While a study of the cross-over between these
three regimes requires the full numerics,  we show here that by
identifying the appropriate time-scales it is possible to derive
vastly simpler equations for each of the three regimes.  The
simplified equations allow a clear physical interpretation, are
easily solved, and are in good agreement with the full numerics in
their respective domains of validity.

\end{abstract}



\section{Introduction}
A generic example of a nanoelectromechanical (NEMS) device is
given by the {\it charge shuttle} (originally proposed by Gorelik
{\it et al.}~\cite{gor-prl-98}): a movable single-electron device,
working in the Coulomb blockade regime, which can exhibit regular
charge transport, where one electron within each mechanical
oscillation cycle is transported from the source to the drain --
see Figure 1 for a schematic illustration (see also
Ref.\cite{sch-njp-02}, which contains an illustrative computer
animation).

\begin{figure}[h]
 \begin{center}
  \includegraphics[angle=0,width=.7\textwidth]{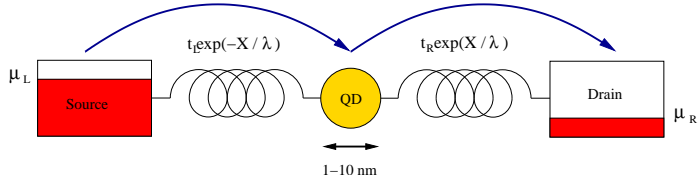}
  \caption{\small  \textit{Schematic representation of the single-dot shuttle: electrons tunnel
        from the left lead at chemical potential ($\mu_L$) to the quantum
        dot and eventually to the right lead at lower chemical potential $\mu_R$.
        The position dependent tunneling amplitudes are indicated.
        $X$ is the displacement from the equilibrium position.
        The springs represent the harmonic potential in which the central dot
        can move.}}
  \label{fig:SdS_model}
 \end{center}
\end{figure}
%

The device shown in Fig.~\ref{fig:SdS_model} can exhibit a number
of different charge transport mechanisms, to be discussed below,
and transitions between the various regimes can be induced by
varying a control parameter, such as the applied bias, or the
mechanical damping. Since its introduction, the charge shuttle has
inspired a large number of theoretical papers (see, {\it e.g.},
Refs.
\cite{arm-prb-02,nov-prl-03,fed-prl-04,pis-prb-04,nov-prl-04,fli-prb-04,fli-epl-05,pis-prl-05,fed-prl-05}).
To the best of our knowledge, a clear-cut experimental
demonstration of a shuttling transition  is not yet available,
though significant progress has been made, such as the {\it
driven} shuttle of Erbe {\it et al.}~\cite{erb-prl-01}, or the
nanopillars studied by Scheible and Blick \cite{sch-apl-04}. In
the present paper, we study the quantum shuttle, {\it i.e.}\ a
device where also the mechanical motion needs to be quantized (the
physical condition for this to happen is $\lambda\simeq x_{zp}$,
where $\lambda$ is the tunneling length describing the exponential
decay of the wave-functions into vacuum, and $x_{zp}$ is the
quantum-mechanical zero-point amplitude).  In contrast to many of
the earlier theoretical papers on quantum shuttles
\cite{arm-prb-02,nov-prl-03,nov-prl-04,fli-prb-04}, where
extensive numerical calculations were employed, the main aim here
is to develop simplified models which allow significant analytic
progress, and hence lead to a transparent physical interpretation.
From the previous numerical studies we know that there are (at
least) three well-defined transport regimes for a quantum
nanomechanical device: (i) the tunneling regime, (ii) the shuttle
regime, and (iii) the coexistence regime.  As we show in
subsequent sections, each of these regimes is characterized by
certain inequalities governing the various time-scales, and a
systematic exploitation of these inequalities allows us then to
develop the aforementioned simplified models. In all three cases
we will compare the predictions of the simplified models to the
ones obtained with the full numerics. While in most cases the
comparison is very satisfactory, we do not always observe
quantitative agreement; these discrepancies are analyzed and
directions for future work are indicated.

The paper is organized as follows.  In Sections 2, 3 and 4 we
briefly introduce the microscopic model for the quantum shuttle,
introduce the Klein-Kramers equations for the Wigner functions,
and summarize the phenomenology extracted from previous numerical
studies, respectively.  Section 5 contains the main results of
this paper, \emph{i.e.}, the derivations and analysis of the
simplified models for the three transport regimes.  We end the
paper with a short conclusion of the main results.

%
\section{The Single-Dot Quantum Shuttle}
The Single-Dot Quantum Shuttle (SDQS) consists of a movable
quantum dot (QD) suspended between source and drain leads (see
Fig.~\ref{fig:SdS_model}). The center of mass of the QD is
confined to a potential that, at least for small displacements
from its equilibrium position, can be considered harmonic. Due to
its small geometric size, the QD has a very small capacitance and
thus a charging energy that may exceed the thermal energy $k_B T$
(approaching room temperature in the most recent realizations
\cite{sch-apl-04}). For this reason we assume that only one excess
electron can occupy the device (Coulomb blockade) and we describe
the electronic state of the central dot as a two-level system
(empty/charged). Electrons can tunnel between leads and dot with
tunneling amplitudes which depend exponentially on the position of
the central island due to the decreasing/increasing overlapping of
the electronic wave functions. The Hamiltonian of the model reads:

\begin{equation}    H =H_{\rm sys}+H_{\rm leads}+H_{\rm bath}
      +H_{\rm tun}+H_{\rm int}
\label{eq:SdS-Ham0}
\end{equation}
where

\begin{equation}
\eqalign{
&H_{\rm sys} =\frac{\hat{p}^2}{2 m}
               +\frac{1}{2}m \w^2 \hat{x}^2
               +(\eps_1- e\mathcal{E}
               \hat{x})c_1^{\dag}c_1^{\phd}\\
&H_{\rm leads} = \sum_{k}(\eps_{l_k}
               c^{\dagger}_{l_k}c^{\phd}_{l_k}
               +\eps_{r_k}
               c^{\dagger}_{r_k}c^{\phd}_{r_k})\\
&H_{\rm tun} = \sum_{k}[T_{l}(\hat{x}) c^{\dagger}_{l_k}c_1^{\phd}
+
                         T_{r}(\hat{x}) c^{\dagger}_{r_k}c_1^{\phd}] + h.c.\\
 &H_{\rm bath} + H_{\rm int }= {\rm generic \; heat \; bath}}
\label{eq:SdS-Ham}
\end{equation}
The hat over the position and momentum ($\hat{x},\hat{p}$) of the
dot indicates that they are operators since the mechanical degree
of freedom is quantized. Using the language of quantum optics we
call the movable grain alone the \emph{system}. This is then
coupled to two electric \emph{baths} (the leads) and a generic
heat bath. The system is described by a single electronic level of
energy $\eps_1$ and a harmonic oscillator of mass $m$ and
frequency $\w$. When the dot is charged the electrostatic force
($e \mathcal{E} $) acts on the grain and gives the
\emph{electrical influence} on the mechanical dynamics. The
electric field $\mathcal{E}$ is generated by the voltage drop
between left and right lead. In our model, though, it is kept as
an external parameter, also in view of the fact that we will
always assume the potential drop to be much larger than any other
energy scale of the system (with the only exception  of the
charging energy of the dot).

The leads are Fermi seas kept at two different chemical potentials
($\mu_L$ and $\mu_R$) by the external applied voltage ($\D V =
(\mu_L - \mu_R)/e$ ) and all the energy levels of the system lie
well inside the bias window. The oscillator is immersed into a
dissipative environment that we model as a collection of bosons
coupled to the oscillator by a weak bilinear interaction:
\begin{equation}
\eqalign{
 H_{\rm bath} &= \sum_{\vet{q}} \hbar \w_{\vet{q}}{d_{\vet{q}}}^{\dagger} d_{\vet{q}}\\
 H_{\rm int}  &= \sum_{\vet{q}} \hbar g \sqrt{\frac{2m\w}{\hbar}} \hat{x}({d_{\vet{q}}} +
{d_{\vet{q}}}^{\dagger})}
\end{equation}
where the operator $d_{\vet{q}}^{\dagger}$ creates  a bath boson
with wave number $\vet{q}$. The damping rate is given by:
\begin{equation} \gamma(\w) = 2 \pi g^2 D(\w)
\end{equation}
where $D(\w)$ is the density of states for the bosonic bath at the
frequency of the system oscillator. A bath that generates a
frequency independent $\gamma$ is called Ohmic.

The coupling to the electric baths is introduced with the
tunneling Hamiltonian $H_{\rm tun}$. The tunneling amplitudes
$T_{l}(\hat{x})$ and $T_{r}(\hat{x})$ depend exponentially on the
position operator $\hat{x}$ and represent the \emph{mechanical
feedback} on the electrical dynamics:

\begin{equation} T_{l,r}(\hat{x})=t_{l,r}\exp(\mp\hat{x}/\lambda)
\end{equation}
where $\lambda$ is the tunneling length. The tunneling
rates from and to the leads ($\bar{\Gamma}_{L,R}$) can be expressed
in terms of the amplitudes:

\begin{equation}  \bar{\Gamma}_{L,R}=\langle \G_{L,R}(\hat{x})\rangle
=\left \langle \frac{2 \pi}{\hbar}D_{L,R}\exp\left(\mp \frac{2 \hat{x}}{\lambda}\right)
 |t_{l,r}|^2 \right\rangle
\end{equation}
where $D_{L,R}$ are the densities of states of the left and right
lead respectively and the average is taken with respect to the quantum
state of the oscillator.

The model has three relevant time scales: the period of the
oscillator $2\pi/\w$, the inverse of the damping rate $1/\gamma$
and the average injection/ejection time $1/\bar{\Gamma}_{L,R}$. It
is possible also to identify three important length scales: the
zero point uncertainty $x_{zp} =\sqrt{\frac{\hbar}{2m\w}}$, the
tunneling length $\lambda$ and the displaced oscillator
equilibrium position $d=\frac{e\mathcal{E}}{m \w^2}$. The ratios
between the  time scales and the ratios between length scales
distinguish the different operating regimes of the SDQS.
%


\section{Klein-Kramers Equation}

The shuttle dynamics has an appealing simple classical
interpretation: the name ``shuttle" suggests the idea of
sequential and periodical loading, mechanical transport and
unloading of electrons between a source and a drain lead.
Motivated by the possibility of observing signatures of quantum
dynamics of the mechanical degree of freedom for a nanoscale
shuttle, we decided, following the suggestion of Armour and
MacKinnon \cite{arm-prb-02}, to explore a system with a quantized
oscillator. We express our results in terms of the Wigner function
because in this way we can simultaneously keep the intuitive
phase-space picture and handle the quantum-classical
correspondence \cite{nov-prl-03}.

The phase space of the shuttle device is spanned by  the  triplet
charge -- position -- momentum. Correspondingly, the Wigner
function is constructed from the reduced density matrix $\s_{ii}$
($i=0,1$ indicates the empty and charged states respectively):

\begin{equation}
 W_{ii}(q,p,t) =
 \frac{1}{2 \pi \hbar}
 \int_{-\infty}^{+\infty}\!\!\!d \xi
 \left\langle q-\frac{\xi}{2}\right| \s_{ii}(t) \left|q+\frac{\xi}{2}\right\rangle
 \exp\left(\frac{ip\xi}{\hbar}\right)
 \label{eq:WF}
\end{equation}
where the reduced density matrix $\sigma$ is defined as the trace
over the mechanical and thermal baths of the full density matrix:

\begin{equation}
 \sigma = {\rm Tr}_{\rm B}\{\rho\}
\end{equation}

The dynamics of the shuttle device is then completely described by
the equation of motion for the Wigner distribution
\cite{fed-prl-04,andrea}:

\begin{equation}
\eqalign{
 \fl \phantom{abba}\frac{\partial W_{00}}{\partial t} =&
\left[m \w^2 q\frac{\partial}{\partial p}
 -\frac{p}{m}\frac{\partial}{\partial q}
 +\g \frac{\partial}{\partial p}p
 +\g m \hbar \w \left(n_B + \frac{1}{2} \right)
 \frac{\partial^2}{\partial p^2}\right]W_{00}
 \\
 \fl &+\G_{R}e^{ 2q/\l} W_{11}
 -\G_{L}e^{- 2q/\l}\sum_{n=0}^{\infty}
 \frac{(-1)^n}{(2n)!}
 \left(\frac{\hbar}{\l}\right)^{2n}
 \frac{\partial^{2n}W_{00}}{\partial p^{2n}}
 \\
\fl \phantom{abba}\frac{\partial W_{11}}{\partial t} =&
 \left[m
 \w^2(q-d)\frac{\partial }{\partial p}
 -\frac{p}{m}\frac{\partial}{\partial q}
 +\g \frac{\partial}{\partial p}p
 +\g m \hbar \w \left(n_B + \frac{1}{2} \right)
 \frac{\partial^2}{\partial p^2}\right]W_{11}
 \\
 \fl &+\G_{L}e^{- 2q/\l} W_{00}
 -\G_{R}e^{2q/\l}\sum_{n=0}^{\infty}
 \frac{(-1)^n}{(2n)!}
 \left(\frac{\hbar}{\l}\right)^{2n}
 \frac{\partial^{2n}W_{11}}{\partial p^{2n}}}
 \label{eq:KleinKramers}
\end{equation}
where $(q,p)$ are the position and momentum coordinates of the
mechanical phase space and $n_B$ is the Bose distribution
calculated at the natural frequency of the harmonic oscillator.
Only the diagonal charge states enter the Klein-Kramers equations
\eref{eq:KleinKramers}: the off-diagonal charge states vanish
given the incoherence of the leads and are thus excluded from the
dynamics.

We distinguish in Eqs.~\eref{eq:KleinKramers}  contributions of
different physical origin: the coherent terms that govern the
dynamics of the (shifted) harmonic oscillator, the dissipative
terms proportional to the mechanical damping constant $\gamma$
and, finally the driving terms proportional to the bare tunneling
rates $\Gamma_{L,R}$.

The ability of the formalism to treat the quantum-classical
correspondence is explicit in Eqs.~\eref{eq:KleinKramers}: given a
length, a mass, and a time scale for the system we can rescale the
phase space coordinates and an expansion in $\hbar/S_{\rm sys}$
will appear where $S_{\rm sys}$ is the typical action of the
system. Classical systems have a large action $S_{\rm sys} \gg
\hbar$ and only the first term ($\hbar/S_{\rm sys} \to 0$) in the
expansion is relevant. In the opposite limit $S_{\rm sys} \approx
\hbar$ the full expansion should be considered.

\section{Phenomenology}

The stationary solution of the Klein-Kramers equations for the
Wigner distributions \eref{eq:KleinKramers} describes the average
long time behavior of the shuttling device. Information about the
different long time operating regimes can be extracted from the
distribution itself or from the experimentally accessible
stationary current and zero frequency current noise.

\begin{figure}[h]
 \begin{center}
  \includegraphics[width=.6\textwidth]{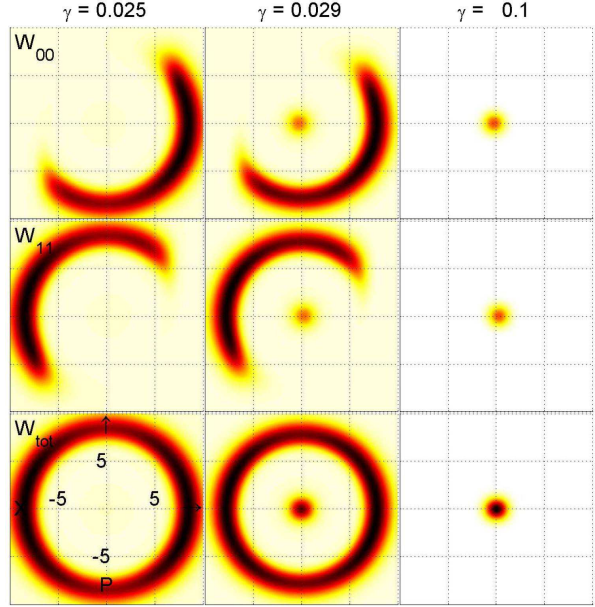}
  \caption{\small
  \textit{Charge resolved Wigner function distributions for different
  mechanical damping rates (horizontal axis: coordinate in units of
  $x_0=\sqrt{\hbar/m\w}$; vertical axis: momentum in $\hbar/x_0$).
  The rows represent from top to bottom the  empty $(W_{00})$, charged
  $(W_{11})$ and total $(W_{\rm tot} = W_{00}+W_{11})$ Wigner distribution
  respectively. The columns represent  from left to right the shuttling
  $(\gamma=0.025\,\w)$, coexistence $(\gamma=0.029\,\w)$ and tunneling
  regime $(\gamma=0.1\,\w)$  respectively. The Figure is partially reproduced from
  \cite{nov-prl-04}.}}
  \label{fig:WF}
 \end{center}
\end{figure}

The mechanical damping rate $\gamma$ is the control parameter of
our analysis. At high damping rates the total Winger distribution
is concentrated around the origin of the phase space and
represents the harmonic oscillator in its ground state. While
reducing the mechanical damping a ring  develops and, after a
short coexistence, the central ``dot'' eventually disappears
(Figure \ref{fig:WF}). The ring is the noisy representation of the
low damping limit cycle trajectory (shuttling) that develops from
the high damping equilibrium position (tunneling). Equilibrium and
limit cycle dynamics coexist in the intermediate damping bistable
configuration where the system randomly switches between tunneling
and shuttling regimes. The charge resolved Winger distributions
$W_{00}$ and $W_{11}$ also reveal the charge-position (momentum)
correlation typical of the shuttling regime: for negative
displacements and positive momentum (\emph{i.e.}\ leaving the
source lead) the dot is prevalently charged while it is empty for
positive displacements and negative momentum (coming from the
drain lead).

\begin{figure}[h]
 \begin{center}
  \includegraphics[angle=-90, width=.48\textwidth]{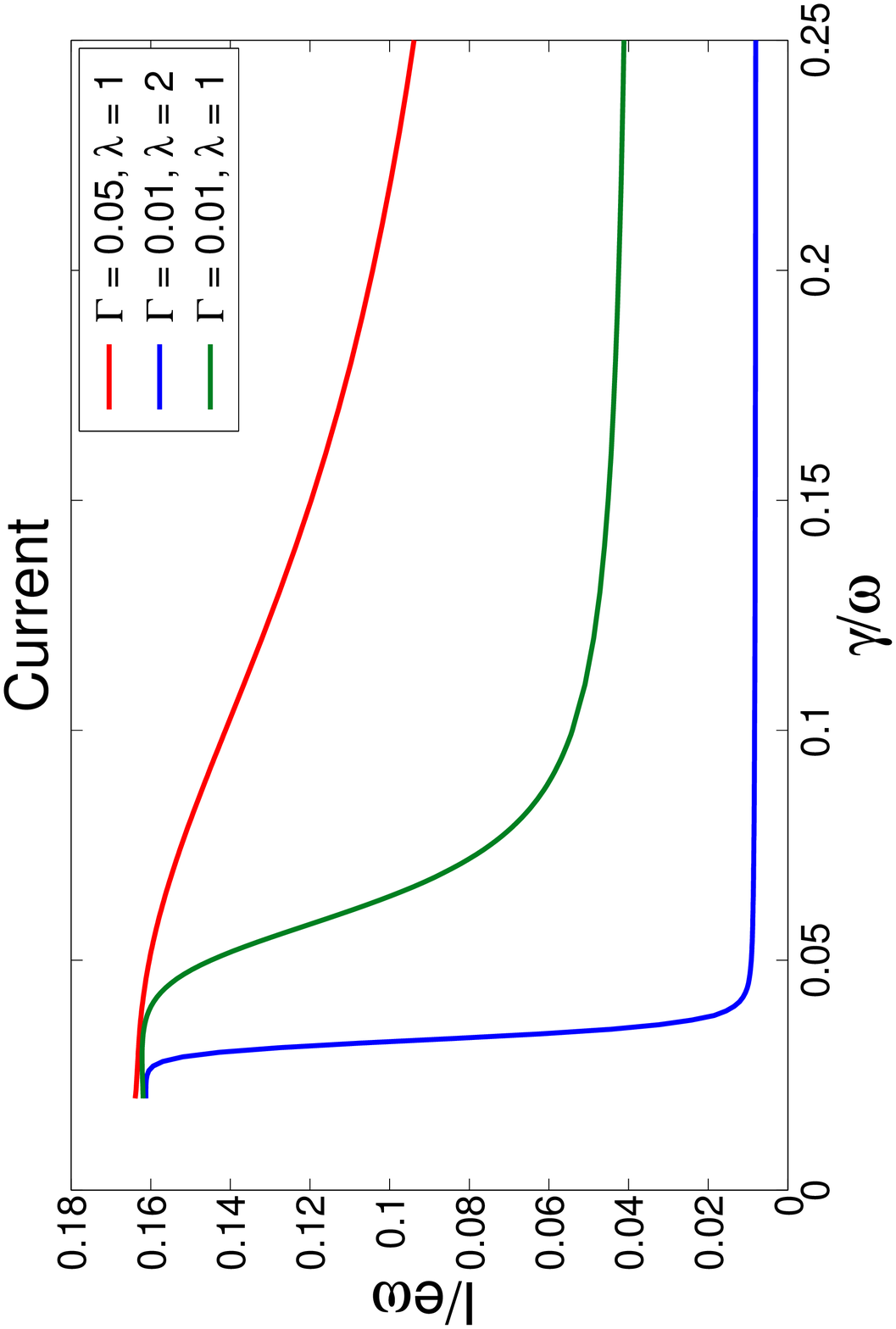}
  \includegraphics[angle=-90, width=.48\textwidth]{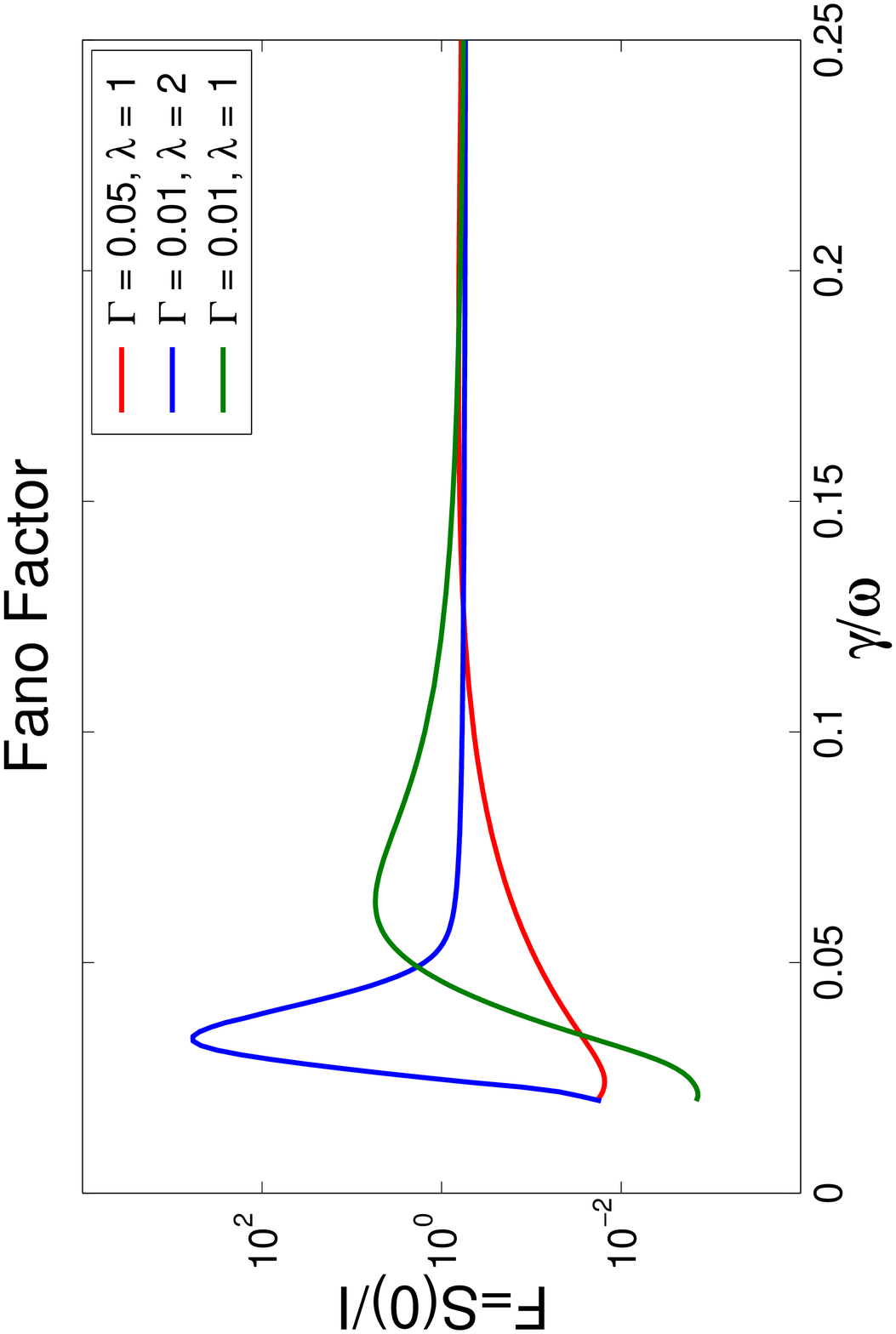}
  \caption{\small  \textit{Left panel - Stationary current for the SDQS vs.~damping $\gamma$.
  The mechanical dissipation rate $\gamma$ and the electrical rate $\G = \G_L = \G_R$ are
  given in units of the mechanical frequency $\w$,
  the tunneling length $\l$ in terms of $x_0 = \sqrt{\hbar/m\w}$. The
  other parameters are $d = 0.5\,x_0$ and $T=0$. The current
  saturates in the shuttling (low damping) regime to one electron
  per cycle independently from the parameters while is
  substantially proportional to the bare electrical rate $\G = \G_L =
  \G_R$ in the tunneling regime (high damping).
  Right panel - Fano factor for the SDQS vs.~damping $\gamma$. The curves correspond to the same
  parameters of the left panel.
  The very low noise in the shuttling  (low damping) regime is a sign of ordered transport.
  The huge super-poissonian Fano factors correspond to the onset of the coexistence regime.
  The Figure is taken from \cite{nov-prl-04}.}}
  \label{fig:CurrNoiseSDQS}
 \end{center}
\end{figure}

Also the stationary current and the current noise (expressed in
terms of the Fano factor) show distinctive features for the
different operating regimes. At high damping rates the shuttling
device behaves essentially like the familiar double-barrier system
since the dot is (almost) static and far from both electrodes. The
current is determined essentially by the bare tunneling rates
$\Gamma_{L,R}$ and the Fano factor differs only slightly  from the
values found for resonant tunneling devices ($F=1/2$ for a
symmetric device). At low damping rates the current saturates at
one electron per mechanical cycle (corresponding to current
$I/e\omega=1/2\pi$) since the electrons are shuttled one by one
from the source to the drain lead by the oscillating dot while the
extremely sub-poissonian Fano factor reveals the deterministic
character of this electron transport regime. The fingerprint of
the coexistence regime, at intermediate damping rates, is a
substantial enhancement of the Fano factor. The current
interpolates smoothly between the shuttling and tunneling limiting
values (Figure \ref{fig:CurrNoiseSDQS}).

The cross-over damping rate is determined by the effective
tunneling rates of the electrons. We get the following physical
picture: every time an electron jumps on the movable grain the
grain is subject to the electrostatic force $e\mathcal{E}$ that
accelerates it towards the drain lead. Energy is pumped into the
mechanical system and the dot starts to oscillate. If the damping
is high compared to the tunneling rates the oscillator dissipates
this energy into the environment before the next tunneling event:
on average the dot remains in its ground state. On the other hand,
for very small damping the relaxation time of the oscillator is
long and multiple ``forcing events" occur before the relaxation
takes place. This continuously drives the oscillator away from
equilibrium and a stationary state is reached only when the energy
pumped per cycle into the system is dissipated during the same
cycle in the environment.

\section{Simplified models} \label{sec:simplified_models}

We qualitatively described in the previous  section three possible
operating regimes for shuttle-devices. The specific separation of
time scales allows us to identify the relevant variables and
describe each regime by a specific simplified model. Models for
the tunneling, shuttling and coexistence regime are analyzed
separately in the three following subsections. We also give a
comparison with the full description in terms of Wigner
distributions, current and current-noise to illustrate how the
models capture the relevant dynamics.

\subsection{Renormalized resonant tunneling}
\label{sec:tunneling}

The electrical dynamics has the longest time-scale in the
tunneling regime since the mechanical relaxation time (which is
much longer than the oscillation period) is much shorter than the
average injection or ejection time. Because of this time-scale
separation, the observation of the device dynamics would most of
the time show two mechanically frozen states:

\begin{description}
 \item[0.]Empty dot in the ground state
 \item[1.]Charged dot moved to the shifted equilibrium position
 by the constant electrostatic force $e\mathcal{E}$.
\end{description}
We combine this observation with a quantum description  of the
mechanical oscillator and possible thermal noise under the
assumption that the reduced density matrix of the device can be
written in the form:

\begin{equation}
 \eqalign{
  \s_{00}(t) &= p_{00}(t)\s_{\rm th}(0)\\
  \s_{11}(t) &= p_{11}(t)\s_{\rm th}(e\mathcal{E})}
 \label{eq:tun-ansatz}
\end{equation}
where
\begin{equation} \s_{\rm th}(\mathcal{F}) =
 \frac{e^{-\b(H_{\rm osc} -\mathcal{F} x)}}
 {{\rm Tr}_{\rm osc}\left[e^{-\b(H_{\rm osc} -\mathcal{F} x)}\right]}
\end{equation}
is the thermal density matrix of a harmonic oscillator  subject to
an external force $\mathcal{F}$. The functions $p_{00}(t)$ and
$p_{11}(t)$ represent the probability to find the system
respectively in the state 0 or 1, respectively. The equations of
motion for the probabilities $p_{ii}(t)$ can be derived by
inserting the assumption (\ref{eq:tun-ansatz}) in the definition
\eref{eq:WF} and taking the integral over the mechanical degrees
of freedom in the corresponding Klein-Kramers equations
\eref{eq:KleinKramers}. This results in the rate equations

\begin{equation}
\frac{d}{dt} \left(
\begin{array}{c}
 p_{00}\\
 p_{11}
\end{array}
\right)
 =
\left(
\begin{array}{c}
 \tilde{\G}_R\, p_{11} -\tilde{\G}_L\, p_{00} \\
 \tilde{\G}_L\, p_{00} - \tilde{\G}_R\, p_{11}
\end{array}
\right) \equiv \mathcal{L} \left(
\begin{array}{c}
 p_{00}\\
 p_{11}
\end{array}
\right)
 \label{eq:ME-tun}
\end{equation}
where
\begin{equation}\eqalign{
 \tilde{\G}_L &= \G_L \Tr_{\rm mech}\left\{\s_{\rm
 th}(0)e^{-\frac{2\hat{x}}{\l}}\right\}=
  \G_L \int d q\,d p\, e^{-\frac{2q}{\l}}W_{\rm th}(q,p)\\
 \tilde{\G}_R &= \G_R \Tr_{\rm mech}\left\{\s_{\rm
 th}(e\mathcal{E})e^{\frac{2\hat{x}}{\l}}\right\}=
  \G_R\int d q\,d p\,  e^{\frac{2q}{\l}}W_{\rm th}(q-d,p)}
  \label{eq:Rates1}
\end{equation}
are the renormalized injection and ejection rates and $\Tr_{\rm
mech}$ indicates the trace over the mechanical degrees of freedom
of the device. We have also introduced the Liouvillean operator:

\begin{equation} \mathcal{L} =
\left(%
\begin{array}{cc}
  -\tilde{\G}_L & \tilde{\G}_R\\
   \tilde{\G}_L &-\tilde{\G}_R
\end{array}%
\right) \label{eq:Liouville}
 \end{equation}
The thermal equilibrium Wigner function $W_{\rm th}(q-d,p)$ is
defined as the Wigner representation of the thermal equilibrium
density matrix $\s_{\rm th}(e\mathcal E)=\s_{\rm th}(m\w^2 d)$:

\begin{equation}
W_{\rm th}(q-d,p) = \frac{1}{ 2 \pi m\w\ell^2}
\exp\left\{-\frac{1}{2}\left[\left(\frac{q-d}{\ell}\right)^2 +
                     \left(\frac{p}{\ell m\w}\right)^2 \right]\right\}
\label{eq:Wth}
\end{equation}
where $\ell = \sqrt{\frac{\hbar}{2m\w}(2n_B +1)}$  reduces to the
zero point uncertainty length $x_{zp}=\sqrt{\frac{\hbar}{2m\w}}$
in the zero temperature limit. In the high temperature limit $k_B
T \gg \hbar \w$, $\ell$ tends  to the thermal length
$\l_{\rm{th}}=\sqrt{k_B T/(m\w^2)}$. Using (\ref{eq:Wth}) in
\eref{eq:Rates1} gives the renormalized rates:

\begin{equation} \eqalign{
 \tilde{\G}_L &= \G_L e^{2\left(\frac{\ell}{\l}\right)^2}\\
 \tilde{\G}_R &= \G_R e^{\frac{2d}{\l} + 2\left(\frac{\ell}{\l}\right)^2}
 }
 \label{eq:Ratesfinal}
\end{equation}
Equations (\ref{eq:ME-tun}) describe the dynamics of a resonant
tunneling device. All the effects of the movable grain are
contained in the effective rates $\tilde{\G}_L,\tilde{\G}_R$. As
expected the ejection rate  is modified by the ``classical" shift
$d$ of the equilibrium position due to the electrostatic force on
the charged dot. Note that both rates are also \emph{enhanced} by
the fuzziness in the position of the oscillator due to thermal and
quantum noise. The relevance of this correction is given by the
ratio between $\ell$ and the tunneling length $\l$.

\subsubsection{Phase-space distribution}

The phase space distribution for the stationary state of the
simplified model for the tunneling regime is built on the Wigner
representation of the thermal density matrix $\s_{\rm th}$ and the
stationary solution of the system (\ref{eq:ME-tun}) for the
occupation $p_{ii}$ of the electromechanical states $i$:
\begin{equation}\eqalign{
 W_{00}^{\rm stat}(q,p) &=
  \frac{\tilde{\G}_R}{\tilde{\G}_L + \tilde{\G}_R} W_{\rm th}(q,p)\\
 W_{11}^{\rm stat}(q,p) &=
  \frac{\tilde{\G}_L}{\tilde{\G}_L + \tilde{\G}_R} W_{\rm th}(q-d,p)
}
\end{equation}
%
%
\begin{figure}[ht]
 \begin{center}
  \includegraphics[angle=-90,width=.8\textwidth]{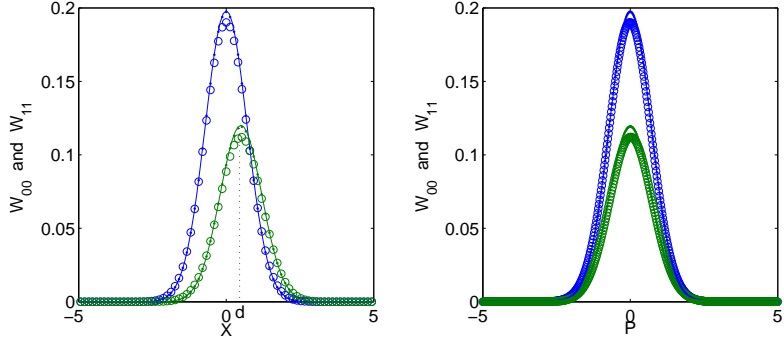}
  \caption{\small  \textit{Comparison between the numerical and the analytical results
  for the Wigner distribution functions. The coordinate (left) or momentum (right) cuts always
  cross the maximum of the distribution. The green (blue) circles are numerical results
  in the charged (empty) dot configuration, and the full lines
  represent the analytical calculations. The parameters are:
   $\G_L = \G_R = 0.01\,\w$, $\g=0.25\,\w$, $d = 0.5\,x_0$, $\l = 2\,x_0$, $T=0$.
  We also plotted with dots the numerical results for $\G = 0.001\,\omega$.}
  \label{fig:Cutstun}}
 \end{center}
\end{figure}
%
The stationary distribution of the tunneling model is  determined
by the length $\ell$ and associated momentum $m\w\ell$, the
equilibrium position shift $d$, the tunneling length $\l$ and the
ratio between left and right bare electrical rates
$\Gamma_L/\Gamma_R$. The mechanical relaxation rate $\gamma$ drops
out from the solution and only sets the range of applicability of
the simplified model.

\begin{figure}[ht]
 \begin{center}
  \includegraphics[angle=-90,width=.8\textwidth]{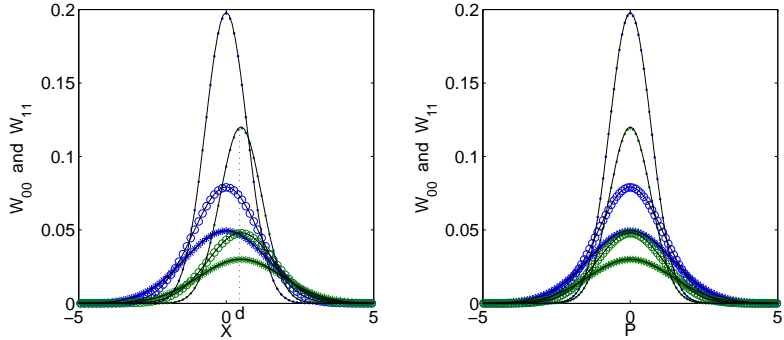}
  \caption{\small  \textit{Tunneling Wigner distributions as a function of the temperature.
  The  relevant parameters are: $\gamma = 0.25\,\w$, $\G = 0.001\,\omega$, $n_B = 0,0.75,1.5$ respectively
  represented by dots, circles and asterisks. Full lines are the analytical results.}
  \label{fig:CutstunTemp}}
 \end{center}
\end{figure}

In Figures \ref{fig:Cutstun} and \ref{fig:CutstunTemp} we compare
the Wigner functions calculated both analytically and numerically
in the tunneling regime. They show in general a good agreement
(Figure \ref{fig:Cutstun}). The matching is further improved when
reducing the bare injection rate $\G_{L,R} = 0.001\omega$ thus
enlarging the time scale separation $\G_{L,R} \ll \g$ typical of
the tunneling regime. The temperature dependence of the stationary
Wigner function distribution (Figure\ \ref{fig:CutstunTemp})
verifies the scaling given by the temperature dependent length
$\ell$.
%

\subsubsection{Current}

Since the effect of the oscillator degree of freedom is entirely
included in the renormalized rates,  the system can be treated
formally as a static quantum dot.
The time-dependent currents thus read:
\begin{equation}\eqalign{
 I_R(t) &= \tilde{\G}_R p_{11}(t)\\
 I_L(t) &= \tilde{\G}_L p_{00}(t)
}
\end{equation}
In the stationary limit they coincide:
\begin{equation}
I^{\rm stat}= \tilde{\G}_R p^{\rm stat}_{11} = \tilde{\G}_L p^{\rm
stat}_{00}  = \frac{\tilde{\G}_R
\tilde{\G}_L}{\tilde{\G}_L+\tilde{\G}_R} \label{eq:Currentun}
\end{equation}
We show in Figure \ref{fig:Currentun} the current calculated
numerically and the asymptotic value of the tunneling regime given
by Eq.~\eref{eq:Currentun}.

\subsubsection{Current-noise}

\begin{figure}
 \begin{center}
  \includegraphics[angle=-90,width=.45\textwidth]{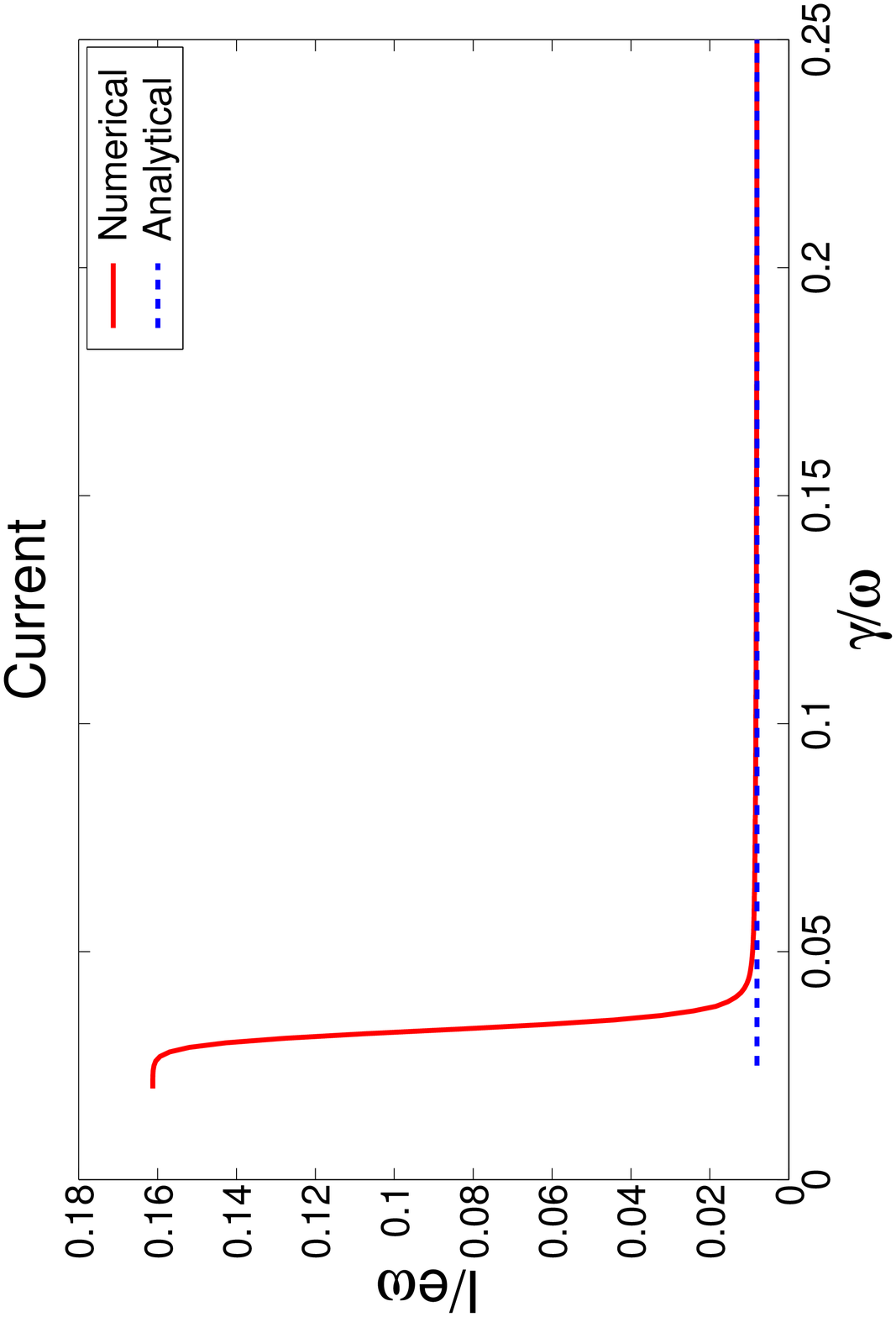}
  \includegraphics[angle=-90,width=.45\textwidth]{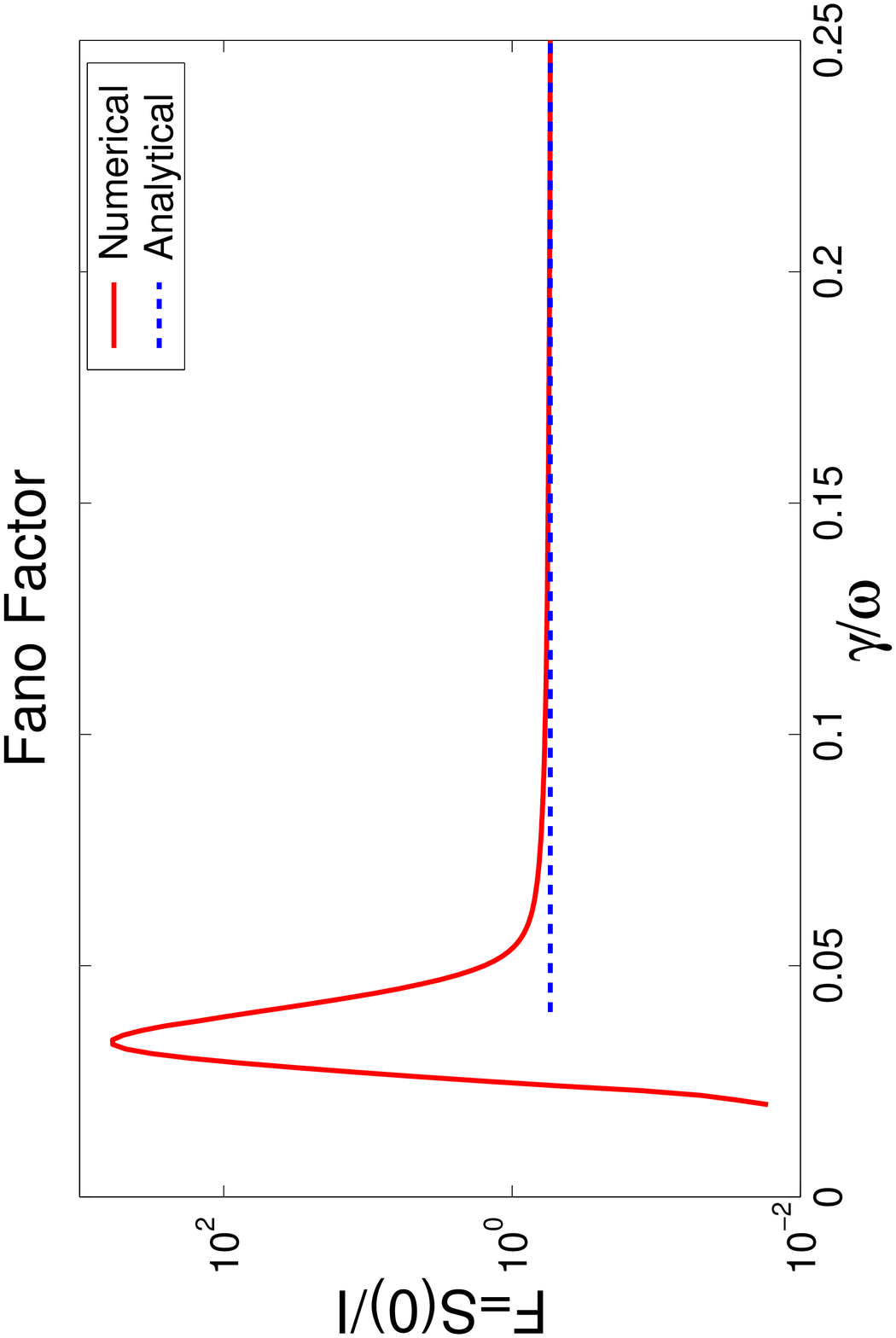}
  \caption{\small  \textit{Left panel - Current as a function of the damping for the SDQS.
  The asymptotic tunneling limit is indicated. The parameters are:
  $\G_L = \G_R = 0.01\,\w$, $\g=0.25\,\w$, $d = 0.5\,x_0$, $\l = 2\,x_0$, $T=0$.
  Right panel - Current-noise as a function of the damping for the SDQS.
  The asymptotic tunneling limit is indicated. The parameters are
  the same as the ones reported for the current.
  }
   \label{fig:Currentun}}
 \end{center}
\end{figure}

We start the calculation with the MacDonald formula for the zero
frequency current noise \cite{fli-prb-04,andrea,ela-pla-02}:

\begin{equation} S(0) = \lim_{t \to \infty}\frac{d}{dt}
 \Big[\sum_{n = 0}^{\infty} n^2P_n(t) -
 \big(\sum_{n= 0}^{\infty}nP_n(t)\big)^2\Big]
 \label{eq:Macdonald}
\end{equation}
where $P_n(t)$ is the probability that $n$ electrons have been
collected at time $t$ in the right lead. This probability is
connected to the $n$-resolved probabilities $p_{ii}^{(n)}$ of the
two effective states of the tunneling model by the relation:
\begin{equation}
 P_n(t)=p_{00}^{(n)}(t) + p_{11}^{(n)}(t)
\end{equation}
The $n$-resolved probabilities $p_{ii}^{(n)}$ satisfy the equation
of motion:

\begin{equation}
\frac{d}{dt} \left(
\begin{array}{c}
 p^{(n)}_{00}\\
 p^{(n)}_{11}
 \end{array}\right)=
 \left(
\begin{array}{c}
 \tilde{\G}_R\, p^{(n-1)}_{11} -\tilde{\G}_L\, p^{(n)}_{00} \\
 \tilde{\G}_L\, p^{(n)}_{00} - \tilde{\G}_R\, p^{(n)}_{11}
\end{array}
 \right)
\end{equation}
that can be derived by tracing the equation of motion for the
total density matrix $\rho$ over bath states with a fixed number
($n$) of electrons collected in the right lead and finally
integrating over the mechanical degrees of freedom. The evaluation
of the different terms of the current-noise \eref{eq:Macdonald}
can be carried out by introducing the generating functions
$F_{ii}(t;z) = \sum_n p_{ii}^{(n)}(t)z^n$
\cite{nov-prl-04,andrea}. The Fano factor is calculated in terms
of the stationary probabilities $p_{ii}^{\rm stat}$ and the
pseudoinverse of the Liouvillean \eref{eq:Liouville}
$\mathcal{QL}^{-1}\mathcal{Q}$.
\begin{equation} F = 1 - \frac{2}{I^{\rm stat}}
 \left(\begin{array}{cc} 1 & 1 \end{array}\right)\left(%
 \begin{array}{cc}
 0 & \tilde{\G}_R\\
 0 & 0
 \end{array}\right)
 \mathcal{QL}^{-1}\mathcal{Q}
 \left(%
 \begin{array}{cc}
 0 & \tilde{\G}_R\\
 0 & 0
 \end{array}\right)
 \left(%
 \begin{array}{c}
 p_{00}^{\rm stat}\\
 p_{11}^{\rm stat}
 \end{array}\right)
 \label{eq:Fanotun}
\end{equation}
For a detailed evaluation of the formula \eref{eq:Fanotun} we
refer the reader to the section IV B of \cite{jau-preprint-04}.
The resulting Fano factor

\begin{equation}
F = \frac{\tilde{\G}_L^2 + \tilde{\G}_R^2}
          {(\tilde{\G}_L + \tilde{\G}_R)^2}
\end{equation}
assumes the familiar form for a tunneling junction, albeit with
renormalized rates. In Fig.~\ref{fig:Currentun} the value of the
Fano factor given by the above formula is depicted as the
high-damping asymptote of the full calculation.

\subsection{Shuttling: a classical transport regime}
\label{sec:shuttling}

The simplified model for the shuttling dynamics is based on the
observation -- extracted from the full description -- that the
system exhibits in this operating regime extremely low Fano
factors ($F \approx 10^{-2}$): we assume that there is \emph{no
noise at all} in the system. Its state is represented by a point
that moves on a trajectory in the  device phase-space spanned by
position, momentum and charge of the oscillating dot. The charge
on the oscillating dot is a stochastic variable governed by
tunnelling processes, however in the shuttling regime the
tunnelling events are effectively deterministic since they are
highly probable only at specific times (or positions) defined by
the mechanical dynamics.

\subsubsection{Equation of motion for the relevant variables}

We implement the zero noise assumption in the set of coupled
Klein-Kramers equations (\ref{eq:KleinKramers}) in two steps: we
first set $T=0$ and then simplify the equations further by
neglecting all the terms of the $\hbar$ expansion since we assume
the classical action of the oscillator to be much larger than the
Planck constant. We obtain:

\begin{equation}
 \eqalign{
 \frac{\partial W_{00}^{\rm cl}}{\partial \t} =&
 \left[X\frac{\partial}{\partial P} - P\frac{\partial}{\partial X} +
  \frac{\g}{\w} \frac{\partial}{\partial P}P\right]\,W_{00}^{\rm cl}\\
 &-\frac{\G_L}{\w} e^{-2X} W_{00}^{\rm cl} +\frac{\G_R}{\w} e^{2X} W_{11}^{\rm cl}\\
\frac{\partial W_{11}^{\rm cl}}{\partial \t} =&
 \left[\left(X - \frac{d}{\l}\right)\frac{\partial}{\partial P} - P\frac{\partial}{\partial X} +
  \frac{\g}{\w} \frac{\partial}{\partial P}
 P\right]\,W_{11}^{\rm cl}\\
 &-\frac{\G_R}{\w} e^{ 2X} W_{11}^{\rm cl}
  +\frac{\G_L}{\w} e^{-2X} W_{00}^{\rm cl}
} \label{eq:FPshuttling}
\end{equation}
where we have introduced the dimensionless variables:
\begin{equation}
  \t = \w t, \quad X = \frac{q}{\l}, \quad P = \frac{p}{m \w \l}
  \label{eq:nodimension}
\end{equation}
The superscript ``cl" indicates that we are dealing with the
classical limit of the Wigner function because of the complete
elimination of the quantum ``diffusive" terms from the
Klein-Kramers equations.  In this spirit, it is natural to try an
Ansatz for the Wigner functions, in which the position and
momentum dependencies are separable:

\begin{equation}
 \eqalign{
 W_{00}^{\rm cl}(X,P,\t) = p_{00}(\t)\d(X-X^{\rm cl}(\t))\d(P-P^{\rm cl}(\t))\\
 W_{11}^{\rm cl}(X,P,\t) = p_{11}(\t)\d(X-X^{\rm cl}(\t))\d(P-P^{\rm cl}(\t))
 }
 \label{eq:separation}
\end{equation}
where the trace over the system phase-space sets the constraint
$p_{00} + p_{11} = 1$. The variables $X^{\rm cl}$ and $P^{\rm cl}$
represent the position and momentum of the (center of mass) of the
oscillating dot; $p_{11(00)}$ is the probability for the quantum
dot to be charged (empty).

By inserting the Ansatz (\ref{eq:separation}) into equation
(\ref{eq:FPshuttling}) and matching the coefficients of the terms
proportional to $\delta\times\delta$ we obtain the equations of
motion for the charge probabilities $p_{ii}$:

\begin{equation}
 \eqalign{
 \dot{p}_{00} &=-\frac{\G_L}{\w}e^{-2X^{\rm cl}}p_{00}
                +\frac{\G_R}{\w}e^{2X^{\rm cl}}p_{11}\\
 \dot{p}_{11} &= \frac{\G_L}{\w}e^{-2X^{\rm cl}}p_{00}
                -\frac{\G_R}{\w}e^{2X^{\rm cl}}p_{11}
 }
 \label{eq:electrical}
\end{equation}
Matching the coefficients proportional to the distributions
$\delta\times\delta'$ (here $\delta'$ is the derivative of the
delta-function) yields the equations for the mechanical degrees of
freedom:

\begin{equation}
 \eqalign{
 p_{00}\dot{X}^{\rm cl} &= p_{00} P^{\rm cl}\\
 p_{11}\dot{X}^{\rm cl} &= p_{11} P^{\rm cl}\\
 p_{00}\dot{P}^{\rm cl} &= p_{00}(-X^{\rm cl} - \frac{\g}{\w} P^{\rm cl})\\
 p_{11}\dot{P}^{\rm cl} &= p_{11}(-X^{\rm cl} +\frac{d}{\lambda}- \frac{\g}{\w} P^{\rm cl})\\
 }
 \label{eq:mechanical1}
\end{equation}
The equations involving $\dot{P}^{\rm cl}$ have a solution only if
\begin{equation}
  p_{00}p_{11} = 0
  \label{eq:condition}
\end{equation}
combined with the normalization condition $p_{00} + p_{11} = 1$.
Under these conditions the system \eref{eq:mechanical1} is
equivalent to
\begin{equation}
 \eqalign{
 \dot{X}^{\rm cl} &= P^{\rm cl}\\
\dot{P}^{\rm cl} &= -X^{\rm cl} +\frac{d}{\lambda}p_{11}- \frac{\g}{\w} P^{\rm cl}\\
 }
 \label{eq:mechanical}
\end{equation}
The condition \eref{eq:condition} also follows by substituting the
Ansatz \eref{eq:separation} into the equations
\eref{eq:FPshuttling} and by using the equations of motion
\eref{eq:mechanical} and \eref{eq:electrical}. This shows that
\eref{eq:condition} also sets the limits of the validity of the
Ansatz \eref{eq:separation}. However, the only differentiable
solution for \eref{eq:condition} is $p_{00} = 0$ or $p_{11} = 0$
for all times, which is not compatible with the equation of motion
\eref{eq:electrical}. Thus, the Ansatz \eref{eq:separation} does
not yield exact solutions to the original equations
\eref{eq:FPshuttling}.

While an exact solution has not been found, we can still proceed
with the following physical argument. Suppose now that the
switching time between the two allowed states, $p_{11}=1;\,
p_{00}=0$ or  $p_{00}=1;\, p_{11}=0$, is much shorter than the
shortest mechanical time (the oscillator period $T = 2\pi /\w$). A
solution of the system of equations (\ref{eq:mechanical}) and
(\ref{eq:electrical}) with this time scale separation would
satisfy the condition (\ref{eq:condition}) ``almost everywhere"
and, when inserted into (\ref{eq:separation}) would represent a
solution for (\ref{eq:FPshuttling}).

\noindent We rewrite the set of equations (\ref{eq:mechanical})
and (\ref{eq:electrical}) as:

\begin{equation}
 \eqalign{
  \dot{X} &= P\\
  \dot{P} &= -X + d^* Q - \gamma^* P\\
  \dot{Q} &= \G_L^* e^{-2X}(1-Q) -\G_R^* e^{2X}Q
 }
 \label{eq:shuttlingfin}
\end{equation}
where we have dropped the ``cl" superscript,  renamed $p_{11}
\equiv Q$, used the trace condition $p_{00} = 1-p_{11}$, and
defined the rescaled parameters: $d^* = d/\l, \quad \g^* = \g/\w,
\quad \G_{L,R}^* = \G_{L,R}/\w$. In the following section we
analyze the dynamics implied by Eq.~(\ref{eq:shuttlingfin}).

\subsubsection{Stable limit cycles}

\begin{figure}[h]
 \begin{center}
 \includegraphics[angle=0,width=0.5\textwidth]{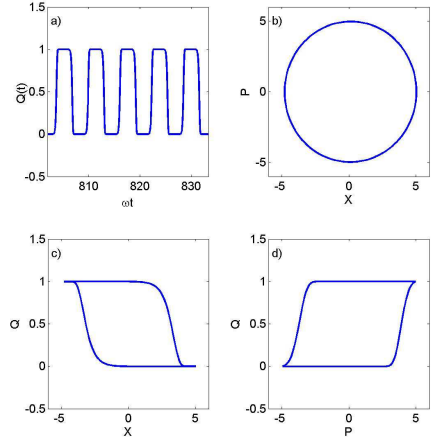}
  \caption{\small  \textit{Different representations of the limit cycle
  solution of the system of differential equations (\ref{eq:shuttlingfin})
  that describes the shuttling regime. For a detailed description see the text.
  $X$ is the coordinate in units of the tunneling length $\l$,
  while $P$ is the momentum in units of $m\w\l$.}
  \label{fig:Limitcycles}}
 \end{center}
\end{figure}

Here we give the results of a numerical solution of
Eq.~(\ref{eq:shuttlingfin}) for different values of the parameters
and different  initial conditions. For the parameter values that
correspond to the fully developed shuttling regime, the system has
a limit cycle solution with the desirable time scale separation we
discussed in the previous section. Figure \ref{fig:Limitcycles}
shows the typical appearance of the limit cycle.

\begin{figure}[h]
 \begin{center}
 \includegraphics[angle=0,width=.5\textwidth]{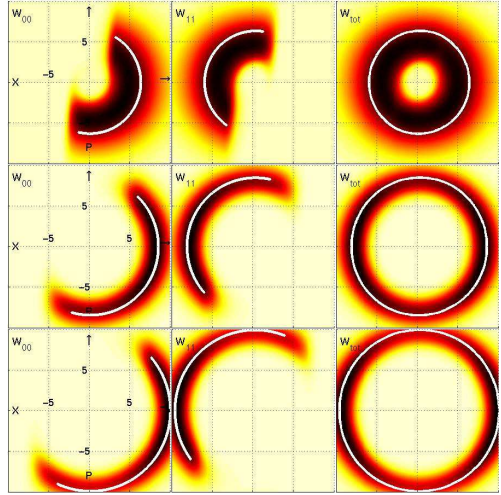}
  \caption{\small  \textit{Correspondence between the Wigner function representation
  and the simplified trajectory limit for the shuttling regime. The white ring is the ($X,P$)
  projection of the limit cycle. The $Q=1$ and $Q=0$ portions of the trajectory are visible in the
  charged and empty dot graphs respectively. The parameters are $\gamma = 0.02\,\w$, $d = 0.5\,x_0$,
  $\G=0.05\,\w$, $\l = x_0$ in the upper row, $\gamma = 0.02\,\w$, $d = 0.5\,x_0$,
  $\G=0.05\,\w$, $\l = 2\,x_0$ in the central row and $\gamma = 0.02\,\w$, $d = 0.5\,x_0$,
  $\G=0.01\,\w$, $\l = 2\,x_0$ in the lower row.}
  \label{fig:rings}}
 \end{center}
\end{figure}

In  Fig.~\ref{fig:Limitcycles}(a) we show the charge $Q(\t)$ as a
function of time. The charge value is jumping periodically from 0
to 1 and back with a period equal to the mechanical period. The
transition itself is almost instantaneous. In panels (b), (c) and
(d) three different projections of the 3D-phase-space trajectory
are reported and the time evolution along them is intended
clockwise. The $X,P$ projection shows the characteristic circular
trajectory of harmonic oscillations. In the $X,Q$ ($P,Q$)
projection the position(momentum)-charge correlation is visible.

The full description of the SDQS in the shuttling  regime has a
phase space visualization in terms of a ring shaped
\emph{stationary} total Wigner distribution function, see
Fig.~\ref{fig:WF}. We can interpret this fuzzy ring as the
probability distribution obtained from many different noisy
realizations of (quasi) limit cycles. The stationary solution for
the Wigner distribution is the result of a diffusive dynamics on
an effective ``Mexican hat" potential that involves both amplitude
and phase of the oscillations. In the noise-free semiclassical
approximation we turn off the diffusive processes and the
point-like state describes in the shuttling regime a single
trajectory with a definite constant amplitude and \emph{periodic}
phase. We expect this trajectory to be the average of the noisy
trajectories represented by the Wigner distribution. In the third
column of Fig.~\ref{fig:rings} the total Wigner function
corresponding to different parameter realization of the shuttling
regime is presented. The white circle is the semiclassical
trajectory. In the first two columns the asymmetric sharing of the
ring between the charged and empty states is also compared with
the corresponding $Q=1$ and $Q=0$ portions of the semiclassical
trajectory.

In the semiclassical description we also have  direct access to
the current as a function of the time. For example the right lead
current reads:

\begin{equation}
 I_R(\t) = Q(\t)\G_R e^{2X(\t)}
\end{equation}
and is also a periodic function with peaks in correspondence to
the unloading processes. The integral of $I_R(\t)$ over one
mechanical period is 1 and represents the number of electron
shuttled per cycle by the oscillating dot, in complete agreement
with the full description.

\subsection{Coexistence: a dichotomous process}

\label{sec:coexistence}

\begin{figure}[h]
 \begin{center}
 \includegraphics[angle=0,width=.6\textwidth]{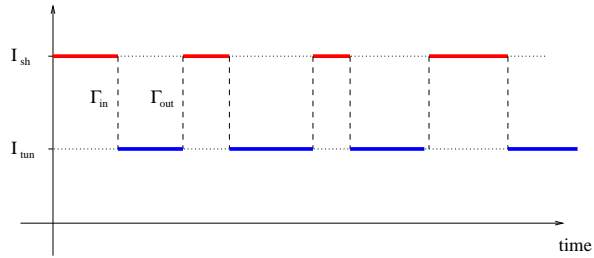}
  \caption{\small  \textit{Schematic representation of the time evolution
  of the current in the dichotomous process between current modes in the SDQS coexistence
  regime. The relevant currents are the
  shuttling ($I_{\rm sh}$) and the tunneling ($I_{\rm tun}$) currents, respectively.
  The switching
  rates $\G_{\rm in}$ and $\G_{\rm out}$ correspond to switching in and out
  of the tunneling mode.}\label{fig:time-scales}}
 \end{center}
\end{figure}

The longest time-scale in the coexistence regime corresponds to
infrequent switching between the shuttling and the tunneling
regime, see Fig.~\ref{fig:time-scales}. The amplitude of the dot
oscillations is the relevant variable that is recording this slow
dynamics. We analyze this particular operating regime of the SDQS
in four steps. (i) We first explore the consequences of the slow
switching in terms of current and current noise. (ii) Next, we
derive the effective bistable potential which controls the
dynamics of the oscillation amplitude. (iii) We then apply
Kramers' theory for escape rates to this effective potential and
calculate the switching rates between the two amplitude metastable
states corresponding to the local minima of the potential. (iv) We
conclude the section by comparing the (semi)analytical results of
the simplified model with the numerical calculations for the full
model.

\subsubsection{Two current modes}\label{sec:dichot}

Let us consider a bistable system with two different modes that we
call for convenience $Shuttling$ (sh) and $Tunneling$ (tun) and
two different currents $I_{\rm sh}$ and $I_{\rm tun}$,
respectively, associated with these modes. The system can switch
between the shuttling and the tunneling mode randomly, but with
definite rates: namely $\G_{\rm in}$ for the process ``shuttling
$\to$ tunneling" and $\G_{\rm out}$ for the opposite, ``tunneling
$\to$ shuttling". We collect this information in the master
equation:

\begin{equation}
 \dot{\vet{P}} = \frac{d}{dt}\left(\begin{array}{c} P_{\rm sh} \\ P_{\rm tun}\\ \end{array}\right) =
 \left(\begin{array}{cc} -\G_{\rm in} & \phantom{-}\G_{\rm out}\\ \phantom{-}\G_{\rm in} & -\G_{\rm out}\\
 \end{array}\right)
 \left(\begin{array}{c} P_{\rm sh} \\ P_{\rm tun} \\ \end{array}\right) = \vet{LP}
 \label{eq:MasterEquation}
\end{equation}
For such a system the average current and the Fano factor read
\cite{andrea,jor-prl-04}:

\begin{equation}
 \eqalign{
 I^{\rm stat} &= \frac{I_{\rm sh} \G_{\rm out} + I_{\rm tun}\G_{\rm in}}{\G_{\rm in}+ \G_{\rm out}}\\
 F &= \frac{S(0)}{I^{\rm stat}} =
  2\frac{(I_{\rm sh}-I_{\rm tun})^2}
        {I_{\rm sh} \G_{\rm out} + I_{\rm tun}\G_{\rm in}}
   \frac{\G_{\rm in}\G_{\rm out}}{(\G_{\rm in}+\G_{\rm out})^2}
 }
 \label{eq:dichCurrFano}
\end{equation}

The framework of the simplified model for the coexistence regime
is given by these formulas. The task is now to identify the two
modes in the dynamics of the shuttle device  and, above all,
calculate the switching rates. This can be done by using the
Kramers' escape rates for a bistable effective potential.

\subsubsection{Effective potential}

\begin{figure}
 \begin{center}
 \includegraphics[angle=0,width=.6\textwidth]{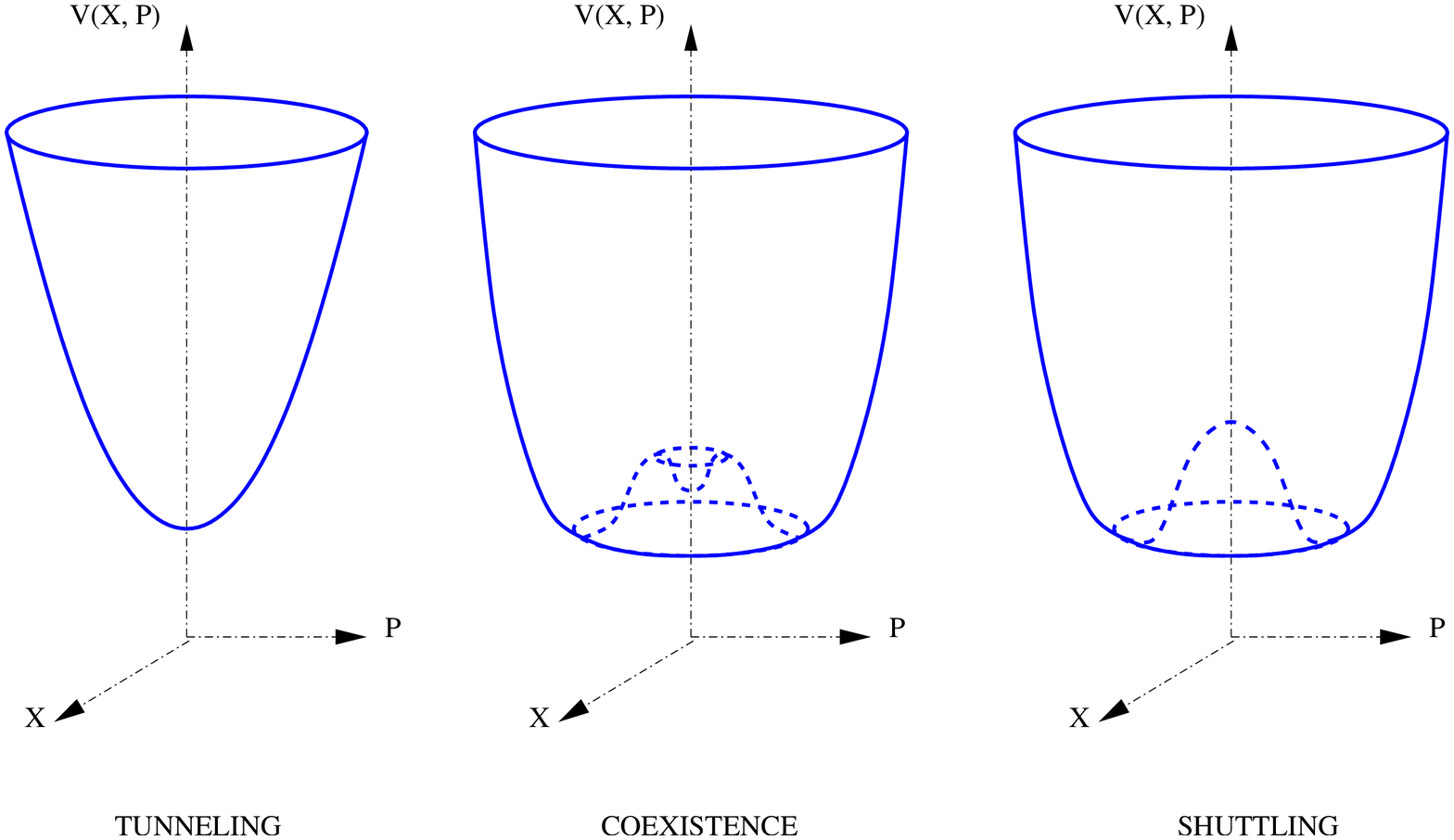}
  \caption{\small  \textit{Schematic representation of the
  effective potentials for the three operating regimes.}
  \label{fig:hats}}
 \end{center}
\end{figure}

The tunneling to shuttling crossover visualized by the total
Wigner function distribution (Fig.~\ref{fig:WF}) can be understood
in terms of an effective stationary potential in the phase space
generated by the non-linear dynamics of the shuttle device. We
show in Fig.~\ref{fig:hats} the three qualitatively different
shapes of the potential surmised from the observation of the
stationary Wigner functions associated with the three operating
regimes. Recently Fedorets {\it et al.}~\cite{fed-prl-04}
initiated the study of the tunneling-shuttling transition in terms
of an effective radial potential. Taking inspiration from their
work we extend the analysis to the slowest \emph{dynamics} in the
device and use quantitatively the idea of the effective potential
for the description of the coexistence regime.

In the process of elimination of the fast variables we start from
the Klein-Kramers equations for the SDQS that we rewrite
symmetrized by shifting the coordinates origin to $d/2$:

\begin{equation}
\fl
 \eqalign{
\frac{\partial W_{00}}{\partial t} =& \left[m \w^2 \left(q +
\frac{d}{2}\right)\frac{\partial}{\partial p}
 -\frac{p}{m}\frac{\partial}{\partial q}
 +\g \frac{\partial}{\partial p}p
 +\g m \hbar \w \left(n_B + \frac{1}{2} \right)
 \frac{\partial^2}{\partial p^2}\right]W_{00}\\
 & +\G_{R}e^{ 2q/\l} W_{11}
 -\G_{L}e^{- 2q/\l}\sum_{n=0}^{\infty}
 \frac{(-1)^n}{(2n)!}
 \left(\frac{\hbar}{\l}\right)^{2n}
 \frac{\partial^{2n}W_{00}}{\partial p^{2n}}\\
\frac{\partial W_{11}}{\partial t} =& \left[m \w^2 \left(q -
\frac{d}{2}\right)\frac{\partial}{\partial p}
 -\frac{p}{m}\frac{\partial}{\partial q}
 +\g \frac{\partial}{\partial p}p
 +\g m \hbar \w \left(n_B + \frac{1}{2} \right)
 \frac{\partial^2}{\partial p^2}\right]W_{11}\\
 & +\G_{L}e^{- 2q/\l} W_{00}
 -\G_{R}e^{2q/\l}\sum_{n=0}^{\infty}
 \frac{(-1)^n}{(2n)!}
 \left(\frac{\hbar}{\l}\right)^{2n}
 \frac{\partial^{2n}W_{11}}{\partial p^{2n}}
 }
 \label{eq:KlKr}
\end{equation}
where the renormalization of the tunneling rates due to the
coordinate shift has been absorbed in a redefinition of the
$\G$'s. The idea is to get rid of the variables that due to their
fast dynamics are not relevant for the description of the
coexistence regime. In equations (\ref{eq:KlKr}) we describe the
electrical state of the dot as empty or charged. We switch to a
new set of variables with the definition:

\begin{equation}
 W_{\pm} = W_{00} \pm W_{11}
\end{equation}
In absence of the harmonic oscillator the state $|+\rangle$ would
be fixed by the trace sum rule and the state $|-\rangle$ would
relax to zero on a time scale fixed by the tunneling rates. We
assume that also in the presence of the mechanical degree of
freedom the relaxation dynamics of the $|-\rangle$ state is much
faster than the one of the $|+\rangle$ state.

In the dimensionless phase space given by the coordinates $X$ and
$P$ of \eref{eq:nodimension} we switch to the polar coordinates
defined by the relations \cite{fed-prl-04}:

\begin{equation}
 X = A \sin \phi \quad P = A \cos \phi
\end{equation}
Since we are interested only in the dynamics of the amplitude in
the phase space (the slowest in the coexistence regime) we
introduce the projector $\mathcal{P}_{\phi}$ that averages over
the phase :

\begin{equation}
 \mathcal{P}_{\phi}[\bullet] = \frac{1}{2\pi}\int_0^{2 \pi} d\phi \bullet
\end{equation}
We also need the orthogonal complement $\mathcal{Q}_{\phi} = 1 -
\mathcal{P}_{\phi}$.  Using these two operators we decompose the
Wigner distribution function into:

\begin{equation}
 W_+ = \mathcal{P}_{\phi} W_+ + \mathcal{Q}_{\phi} W_+ = \bar{W}_+ + \tilde{W}_+
\end{equation}
Finally we make a perturbation expansion of (\ref{eq:KlKr}) in the
small parameters:

\begin{equation}
 \frac{d}{\l} \ll 1,\quad \left(\frac{x_0}{\l}\right)^2 \ll 1, \quad \frac{\g}{\w} \ll 1
 \label{eq:smallparameters}
\end{equation}
These three inequalities correspond to the three physical
assumptions:

\begin{enumerate}

\item The external electrostatic force is a small perturbation of
the harmonic oscillator restoring force in terms of the
sensitivity to displacement of the tunneling rates. This justifies
an oscillator-independent treatment of the tunneling regime.

\item The tunneling length is large compared to the zero point
fluctuations. Since the oscillator dynamics for the shuttling
regime (and then partially also for the coexistence regime)
happens on the scale of the tunneling length, this condition
ensures a quasi-classical behaviour of the harmonic oscillator.

\item The coupling of the oscillator to the thermal bath is weak
and the oscillator dynamics is under-damped.

\end{enumerate}

Using these approximations the Klein-Kramers equations
\eref{eq:KlKr} reduce (for details see, \emph{e.g.},
\cite{fed-prl-04,andrea}) to the form:

\begin{equation}
 \partial_{\t} \bar{W}_+(A,\t)=
 \frac{1}{A}\partial_A A[V'(A) + D(A)\partial_A]\bar{W}_+(A,\t)
 \label{eq:KramersquasiA}
\end{equation}
where $V'(A) = \frac{d}{dA}V(A)$ and $D(A)$ are given functions of
A. Before calculating explicitly the functions $V'$ and $D$ we
explore the consequences of the formulation of the Klein-Kramers
equations (\ref{eq:KlKr}) in the form (\ref{eq:KramersquasiA}).
The stationary solution of the equation (\ref{eq:KramersquasiA})
reads \cite{fed-prl-04}:
\begin{equation}
 \bar{W}_+^{\rm stat}(A) = \frac{1}{\mathcal{Z}}
 \exp\left(-\int_0^A
dA' \frac{V'(A')}{D(A')}\right)\label{eq:FedW}
\end{equation}
where $\mathcal{Z}$ is the normalization that ensures the integral
of the phase-space distribution to be unity: $\int_0^{\infty}
dA'2\pi A' \bar{W}_+^{\rm stat}(A') = 1$. Equation
(\ref{eq:KramersquasiA}) is identical to the Fokker-Planck
equation for a particle in the two-dimensional rotationally
invariant potential $V$ (see Fig.~\ref{fig:hats}) with stochastic
forces described by the (position dependent) diffusion coefficient
$D$. All contributions to the effective potential $V$ and
diffusion coefficient $D$ can be grouped according to the power of
the small parameters that they contain.

\begin{equation}
 \eqalign{
 \fl
 V'(A) &=
  \frac{\g}{\w} \frac{A}{2}
 +\frac{d}{2\l} \a_0(A)
 +\left(\frac{x_0}{\l}\right)^4\a_1(A)
 +\left(\frac{d}{2\l}\right)^2\a_2(A)
 +\frac{\g}{\w}\frac{d}{2\l}\a_3(A)\\
 \fl
 D(A) &=
  \frac{\g}{\w}\left(\frac{x_0}{\l}\right)^2
  \frac{1}{2}\left(n_B + \frac{1}{2}\right)
 +\left(\frac{x_0}{\l}\right)^4\b_1(A)
 +\left(\frac{d}{2\l}\right)^2\b_2(A)
 +\frac{\g}{\w}\frac{d}{2\l}\b_3(A)
 }
\end{equation}
where the $\a$ functions read:
\begin{equation}
 \eqalign{
  \a_0 =&
   \mathcal{P}_{\phi}\cos \phi \hat{G}_0 \G_-\\
  \a_1 =&
   -\frac{1}{4}\mathcal{P}_{\phi} \cos \phi \G_-
   \partial_P (\hat{G}_0 \G_-)\\
  \a_2 =&
   \mathcal{P}_{\phi} \cos \phi \hat{G}_0 \G_-
   \hat{g}_0 \mathcal{Q}_{\phi}
   \partial_P (\hat{G}_0 \G_-)\\
  \a_3 =&
   \mathcal{P}_{\phi}\cos \phi \Big[\hat{G}_0^2 \G_- +
   A \hat{G}_0 \partial_P (\hat{G}_0 \G_-)
   - \frac{A}{2} \sin \phi \partial_P (\hat{G}_0 \G_-)\Big]
 }
\end{equation}
and the $\b$'s can be written as:
\begin{equation}
 \eqalign{
  \b_1 =& \frac{1}{4}
   \mathcal{P}_{\phi}\cos^2\phi \Big[\G_+ - \G_- \hat{G}_0 \G_-\Big]\\
  \b_2 =&
   \mathcal{P}_{\phi} \cos\phi \Big[\hat{G}_0 \cos\phi
   + \hat{G}_0\G_- \hat{g}_0 \mathcal{Q}_{\phi} \cos\phi \hat{G}_0\G_-\Big]\\
  \b_3 =&
   A \mathcal{P}_{\phi}\cos \phi \Big[ \hat{G}_0 \cos^2\phi
   \hat{G}_0 \G_- + \frac{1}{4} \hat{G}_0 \G_- \sin 2\phi
    - \frac{1}{4} \sin 2\phi \hat{G}_0\G_-\Big]\\
   }
\end{equation}
where
\begin{equation}
 \eqalign{
  \Gamma_{\pm}&= \Gamma_L e^{-2A \sin \phi} \pm \Gamma_R e^{2A \sin \phi}\\
 \hat{g}_0 &=(\partial_{\phi})^{-1}\\
 \hat{G}_0 &=(\partial_{\phi} + \G_+)^{-1}
   }
\end{equation}
The $\a$ and $\b$ functions are calculated by isolating in the
Liouvillean for the distribution $\bar{W}_+$ the driving and
diffusive components with generic forms

\begin{equation}
 \frac{1}{A}\partial_A A\{ \a_i(A) \}
 \label{eq:driving}
\end{equation}
and
\begin{equation}
 \frac{1}{A}\partial_A A\{ \b_i(A) \}\partial_A
 \label{eq:diffusive}
\end{equation}
respectively. As an example we give the derivation of the
functions $\a_3$ and $\b_3$. We start by rewriting the equation of
motion \eref{eq:KramersquasiA} for the distribution $\bar{W}_+$ in
the form:
\begin{equation}
 \partial_{\t}\bar{W}_+ = \mathcal{L}[\bar{W}_+] \approx
 (\mathcal{L}^{I} + \mathcal{L}^{II})[\bar{W}_+]
\end{equation}
where we have distinguished the Liouvilleans of first and second
order in the small parameter expansion \eref{eq:smallparameters}.
The contribution $\frac{\g}{\w}\frac{d}{\l}$ of the second order
Liouvillean  $\mathcal{L}^{II}$ reads:
\begin{equation}
\fl
 \mathcal{L}_{\g d} =
   \mathcal{P}_{\phi}\partial_P [\hat{G}_0 \partial_P P \hat{G}_0 \G_- +
   P  \hat{g}_0 \mathcal{Q}_{\phi} \partial_P \hat{G}_0 \G_- +
   \hat{G}_0 \G_- \hat{g}_0 \mathcal{Q}_{\phi} \partial_P P]
\end{equation}
and represents the starting point for the calculation of the
functions $\a_3$ and $\b_3$. We express then the differential
operators $\partial_P$ in polar coordinates and take into account
that the Liouvillean is applied to a function $\bar{W}_+$
independent of the variable $\phi$:

\begin{equation}
\fl
 \eqalign{
 \mathcal{L}_{\g d} = &
 \frac{1}{A} \partial_A A  \mathcal{P}_{\phi} \cos \phi
 \Big[\hat{G}_0^2 \G_- + \hat{G}_0 A \cos \phi \partial_P (\hat{G}_0\G_-)
    + \hat{G}_0 A \cos^2 \phi \hat{G}_0 \G_- \partial_A  \\
 & + A \cos \phi  \hat{g}_0 \mathcal{Q}_{\phi} \partial_P(\hat{G}_0\G_-)
    + A \cos \phi  \hat{g}_0 \mathcal{Q}_{\phi} \cos \phi \hat{G}_0\G_- \partial_A \\
 & +
    \hat{G}_0\G_- \hat{g}_0 \mathcal{Q}_{\phi} A \cos^2 \phi \partial_A \Big]
 }
 \label{eq:quasia3b3}
\end{equation}
Finally, we separate in \eref{eq:quasia3b3} the driving term from
the diffusive contributions and thus identify the functions $\a_3$
and $\b_3$ \footnote{In this step of the derivation we have also
used the projector $\mathcal{P}_{\phi}$ to define a scalar product
$\mathcal{P}_{\phi} f(\phi) g(\phi) \equiv (f,g)$ and the adjoint
relation: $(f, \hat{O}g ) = (\hat{O}^{\dagger}f,g)$.}:

\begin{equation}
\fl
 \eqalign{
 \mathcal{L}_{\g d} = &
  \frac{1}{A} \partial_A A
   \left\{
   \mathcal{P}_{\phi} \cos \phi
   \Big[
     \hat{G}_0^2 \G_- +
   A \hat{G}_0 \partial_P (\hat{G}_0 \G_-)
   - \frac{A}{2} \sin \phi \partial_P (\hat{G}_0 \G_-)
   \Big]
   \right\} +\\
  &\frac{1}{A} \partial_A A
   \left\{
   \mathcal{P}_{\phi} \cos \phi
   \Big[
   \hat{G}_0 \cos^2\phi
   \hat{G}_0 \G_- + \frac{1}{4} \hat{G}_0 \G_- \sin 2\phi
    - \frac{1}{4} \sin 2\phi \hat{G}_0\G_-
   \Big]
   \right\}\partial_A
 }
\end{equation}
Some of these results appear also in the work by Fedorets {\it et
al.}~\cite{fed-prl-04}. Since we have projected out the phase
$\phi$ we are effectively working in a one-dimensional phase space
given by the amplitude $A$. Note however that
Eq.~(\ref{eq:KramersquasiA}) though is \emph{not} as it stands a
Kramers equation for a single variable. This is related to the
fact that also the distribution $\bar{W}_+$ is \emph{not} the
amplitude distribution function, but, so to speak, a cut at fixed
phase of a two dimensional rotationally invariant distribution.
The difference is a geometrical factor $A$. We define the
amplitude probability distribution $\mathcal{W}(A,\t) = A
\bar{W}_+(A,\t)$ and insert this definition in equation
(\ref{eq:KramersquasiA}). We obtain:
\begin{equation}
 \eqalign{
 \partial_{\t} \mathcal{W}(A,\t)
 &= \partial_A A[V'(A) +
 D(A)\partial_A]\frac{1}{A}\mathcal{W}(A,\t)\\
 &= \partial_A [\mathcal{V}'(A) + D(A)\partial_A]\mathcal{W}(A,\t)
 }
 \label{eq:KramersA}
\end{equation}
where we have defined the geometrically corrected potential
\begin{equation}
 \mathcal{V}(A) = V(A) - \int_{A_0}^{A} \!\! \frac{D(A')}{A'} dA'
\end{equation}
which for an amplitude independent diffusion coefficient gives a
corrected potential diverging logarithmically  in the origin. The
lower limit of integration is arbitrary and reflects the arbitrary
constant in the definition of the potential. The equation
(\ref{eq:KramersA}) is the one-dimensional Kramers equation that
constitutes the starting point for the calculation of the
switching rates that characterize the coexistence regime.

\begin{figure}
 \begin{center}
 \includegraphics[angle=0,width=.48\textwidth]{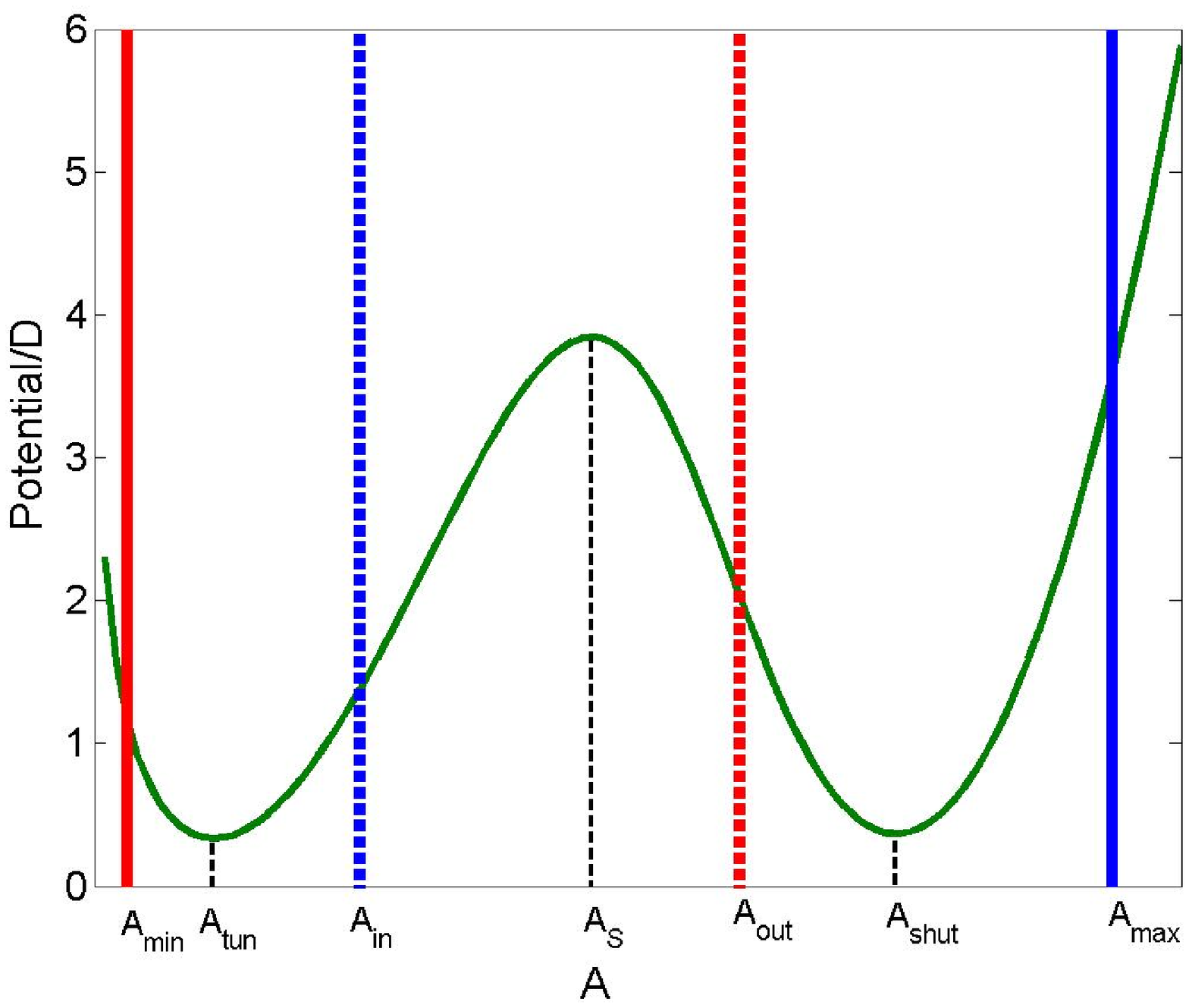}
 \includegraphics[angle=0,width=.48\textwidth]{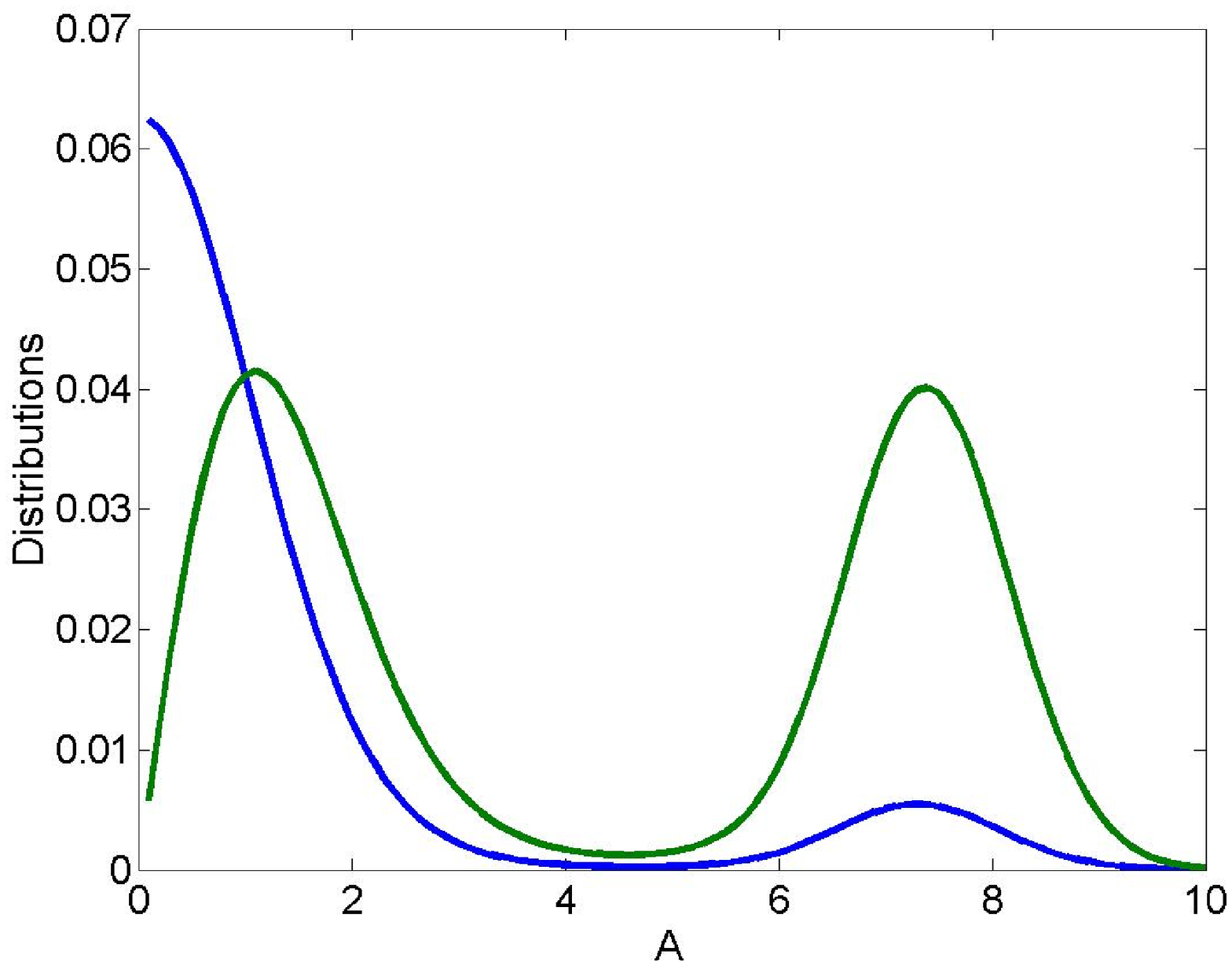}
  \caption{\small  \textit{Left panel -- Bistable effective potential for the SDQS coexistence regime.
  The important amplitudes for the calculation of the rates are indicated. In red (blue) are indicated
  the reflecting (full) and absorbing (dashed) borders for the calculation of the $\G_{\rm out,(in)}$
  escape rate.
  Right panel -- Example of the stationary distribution  $\bar{W}_+^{\rm stat}$ (blue) and the
  amplitude distribution $\mathcal{W}^{\rm stat}$ (green)  for the SDQS in the coexistence regime.
  The tunneling and shuttling states are in both cases well separated.}
  \label{fig:Codisexample}}
 \end{center}
\end{figure}

The effective potential $\mathcal{V}$ that we obtained has, for
parameters that correspond to the coexistence regime, a typical
double-well shape (see \emph{e.g.} the left panel of figure
\ref{fig:Codisexample}). We assume for a while the diffusion
constant to be independent of the amplitude $A$. In this
approximation the stationary solution of the equation
(\ref{eq:KramersA}) reads:

\begin{equation}
 \mathcal{W}^{\rm stat}(A)
 = \frac{1}{\mathcal{Z}} \exp\left(-\frac{\mathcal{V}(A)}{D}\right)
 \label{eq:stationary}
\end{equation}
where $\mathcal{Z}$ is the normalization: $\mathcal{Z} =
\int_0^{\infty} \exp\left(-\frac{\mathcal{V}(A)}{D}\right)dA $.
The probability distribution is concentrated around the minima of
the potential and has a minimum at the potential barrier (see
\emph{e.g} the right panel of Fig. \ref{fig:Codisexample}). If
this potential barrier is high enough (\emph{i.e.}\
$\mathcal{V}_{\rm max} - \mathcal{V}_{\rm min} \gg D$) we clearly
identify two distinct states with definite average amplitude: the
lower amplitude state corresponding to the tunneling regime and
the higher to the shuttling.

The coexistence regime of a SDQS is mapped into a classical model
for a particle moving in a bistable potential $\mathcal{V}$ with
random forces described by the diffusion constant $D$. The
correspondent escape rates from the tunneling to the shuttling
mode ($\G_{\rm out}$) and back ($\G_{\rm in}$) can be calculated
using the standard theory of Mean First Passage Time (MFPT) for a
random variable \cite{risken}:

\begin{equation}
 \eqalign{
 \G_{\rm out}&= D
\left(
 \int_{A_{\rm tun}}^{A_{\rm out}} dB \,
  e^{ \frac{{\mathcal V}(B)}{D}}
 \int_{A_{\rm min}}^{B} dA \,
  e^{-\frac{{\mathcal V}(A)}{D}}
\right)^{-1}\\
 \G_{\rm in}&= D
\left(
 \int_{A_{\rm in}}^{A_{\rm shut}}dB \,
 e^{\frac{{\mathcal V }(B)}{D}}
 \int_{B}^{A_{\rm max}}dA\,
 e^{-\frac{{\mathcal V}(A)}{D}}
 \right)^{-1}\\
 }
 \label{eq:rates}
\end{equation}
where integration limits of equation \eref{eq:rates} are
graphically represented in the left panel of Fig.
\ref{fig:Codisexample}. We can now insert the explicit expression
for the switching rates $\G_{\rm in }$ and $\G_{\rm out}$ in
Eq.~\eref{eq:dichCurrFano} and obtain in this way the current and
Fano factor for the coexistence regime. They represent, together
with the stationary distribution \eref{eq:stationary} the main
result of this section and allow us for a quantitative comparison
between the simplified model and the full description of the
coexistence regime.

\subsubsection{Comparison}

The phase space distribution is the most sensitive object to
compare the model and the full description. One of the basic
procedures adopted in the derivation of the Kramers equation
(\ref{eq:KramersA}) is the expansion to second order in the small
parameters (\ref{eq:smallparameters}). In order to test the
reliability of the model we simplify as much as possible the
description reducing the model to a classical description: namely
taking the zero limit for the parameter
$\left(\frac{x_0}{\l}\right)$. We realize physically this
condition assuming a large temperature and a tunneling length $\l$
of the order of the thermal length $\l_{\rm
th}=\sqrt{\frac{k_BT}{m\w^2}}$. Also the full description is
slightly changed, but not qualitatively: the three regimes are
still clearly present with their characteristics. The numerical
calculation of the stationary density matrix is though based on a
totally different approach. In the quantum regime we used the
Arnoldi iteration scheme for the numerically demanding calculation
of the null vector of the big ($10^4\times 10^4$) matrix
representing the Liouvillean \cite{nov-prl-03,nov-prl-04}.
Problems concerning the convergence of the Arnoldi iteration due
to the delicate issue of preconditioning forced us to abandon this
method in the classical case. We adopted instead the continued
fraction method \cite{risken}.
%
\begin{figure}[h]
 \begin{center}
 \includegraphics[angle=0,width=.5\textwidth]{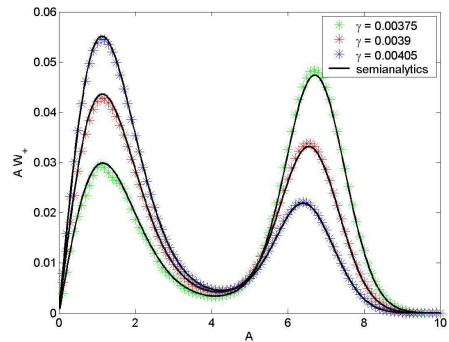}
  \caption{\small
  \textit{Stationary amplitude probability distribution $\mathcal{W}$
  for the SDQS in the coexistence regime. We compare the results obtained from the simplified
  model (full line) and from the full description (asterisks).
  These results are obtained in the classical high temperature regime $k_BT \gg
  \hbar\w$. The amplitude is measured in units of
  $\l_{\rm th} = \sqrt{\frac{k_BT}{m\w^2}}$. The mechanical damping $\g$ in units of the mechanical
  frequency $\w$. The other parameter values are $d = 0.05\,\l_{\rm th}$ and $\G =
 0.015\,\w$, $\lambda = 2\,\lambda_{\rm th}$.}
  \label{fig:CompWF}}
 \end{center}
\end{figure}
%
In Figs.~\ref{fig:CompWF}, \ref{fig:Compcurr} and
\ref{fig:CompNoise} we present the results for the stationary
Wigner function, the current and the Fano factor, respectively, in
the semiclassical approximation and in the full description.
%
\begin{figure}[h]
 \begin{center}
 \includegraphics[angle=0,width=.5\textwidth]{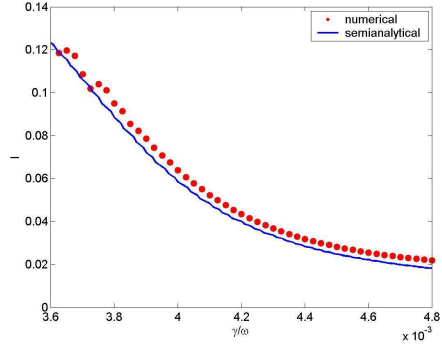}
  \caption{\small  \textit{Current in the coexistence regime of SDQS.
  Comparison between semianalytical and full numerical description.
  For the parameter values see Fig.~\ref{fig:CompWF}.}
  \label{fig:Compcurr}}
 \end{center}
\end{figure}
\begin{figure}[h]
 \begin{center}
 \includegraphics[angle=0,width=.5\textwidth]{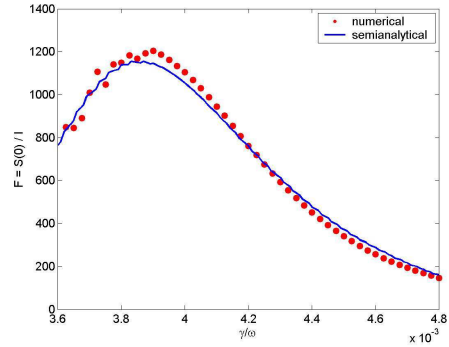}
  \caption{\small  \textit{Fano factor in the coexistence regime of SDQS.
  Comparison between semianalytical and full numerical description.
  For the parameter values see Fig.~\ref{fig:CompWF}.}
  \label{fig:CompNoise}}
 \end{center}
\end{figure}

Deep in the quantum regime the coexistence regime (\emph{e.g.}\
Fig.~\ref{fig:WF} where the amplitude of the shuttling
oscillations is $\approx 7 x_0$) is not captured quantitatively
with the simplified model. Given that the concept of elimination
of the fast dynamics is still valid, we believe that the
discrepancy indicates that the expansion in the small parameters
has not been carried out to sufficiently high order . The
effective potential calculated from a second order expansion still
gives the position of the ring structure with reasonable accuracy
but the overall stationary Wigner function is not fully reproduced
due to an inaccurate diffusion function $D(A)$. One should thus
consider higher order terms in the parameter $(x_0/\l)^2$. A
higher order expansion, however, represents a fundamental problem
since it would produce terms with higher order derivatives with
respect to the amplitude $A$ in the Fokker-Planck equation and,
consequently, a straight forward application of the escape time
theory is no longer possible.

It has nevertheless been demonstrated \cite{fli-epl-05} with the
help of the higher cumulants of the current that the description
of the coexistence regime as a dichotomous process is valid also
deep in the quantum regime ($\l = 1.5\,x_0$), the only necessary
condition being a separation of the ring and dot structures in the
stationary Wigner function distribution.

We observe that the second order expansion for the effective
potential \eref{eq:smallparameters} is essentially converged, and
able to give the correct position of the shuttling ring also in
the quantum regime. From Eq.~\eref{eq:FedW} it is clear that a
strongly amplitude dependent diffusion constant $D(A)$ would
destroy this agreement. We conjecture that a higher order
expansion may lead to an effective renormalization of the
diffusion constant.  To test this idea we used the diffusion
constant as a fitting parameter, and found that the current and
the Fano factor are very accurately reproduced by using a fitted
diffusion constant, with a value approximately twice larger than
the one calculated at zero temperature and in second order in the
small parameters. Clearly, further work is needed to find out
whether the agreement is due to fortuitous coincidence, or if  a
physical justification can be given to this conjecture.  Also, an
investigation of whether the results obtained in the classical
limit are extendable to the quantum limit by a controlled
renormalization of the diffusion constant is called for.

\section{Conclusions}

The specific separation of time scales in the different regimes
allowed us to identify the relevant variables and describe each
regime by a specific \emph{simplified model}. In the tunneling
(high damping) regime the mechanical degree of freedom is almost
frozen and all the features revealed by the Wigner distribution,
the current and the current noise can be reproduced with a
resonant tunneling model with tunneling rates renormalized due to
the movable quantum dot. Most of the features of the shuttling
regime (self-sustained oscillations, charge-position correlation)
are captured by a simple model derived as the zero-noise limit of
the full description. Finally for the coexistence regime we
proposed a dynamical picture in terms of slow dichotomous
switching between the tunneling and shuttling modes. This
interpretation was mostly suggested by the presence in the
stationary  Wigner function distributions of both the
characteristic features of the tunneling and shuttling dynamics
and by a corresponding gigantic peak in the Fano factor. We based
the derivation of the simplified model on the fast variables
elimination from the Klein-Kramers equations for the Wigner
function and a subsequent derivation of an effective bistable
potential for the amplitude of the dot oscillation (the relevant
slow variable in this regime).

The comparison of the results obtained using the simplified models
with the full description in terms of Wigner distributions,
current and current-noise proves that the models, at least in the
limits set by the chosen investigation tools, capture the relevant
features of the shuttle dynamics.

The work of T.~N.\ is a part of the research plan MSM 0021620834
that is financed by the Ministry of Education of the Czech
Republic, and A.~D.\ acknowledges financial support from the
Deutsche Forschungsgemeinschaft within the framework of the
Graduiertenkolleg ``Nichtlinearit\"at und Nichtgleichgewicht in
kondensierter Materie'' (GRK 638).

\section*{References}

\bibliography{shuttle}
\end{document}